\documentclass[lettersize,journal]{IEEEtran}
\usepackage[utf8]{inputenc}
\usepackage[T1]{fontenc}
\usepackage[english]{babel}

\usepackage[numbers]{natbib}
\usepackage{amsmath, amsfonts}
\usepackage{algorithmic}
\usepackage{graphicx}
\usepackage{textcomp}
\usepackage{xcolor}
\usepackage{tcolorbox}
\usepackage{latexsym}
\usepackage[inline]{enumitem}
\usepackage{subcaption}
\usepackage{hyperref}
\usepackage{booktabs}
\usepackage{multirow}
\usepackage{verbatim}
\usepackage{makecell}
\usepackage{enumitem}
\usepackage{listings}
\usepackage{xspace}
\usepackage{color}
\usepackage{rotating}
\usepackage{tabularray}
\usepackage{balance}
\usepackage{float}
\UseTblrLibrary{siunitx}
\definecolor{dkgreen}{rgb}{0,0.6,0}
\definecolor{gray}{rgb}{0.5,0.5,0.5}
\definecolor{mauve}{rgb}{0.58,0,0.82}
\newcommand{\new}[1]{\textcolor{black}{#1}}
\newcommand{\infilling}{infilling\xspace}
\definecolor{darkgreen}{rgb}{0.0, 0.5, 0.0}
\definecolor{darkred}{rgb}{0.5, 0.0, 0.0}
\definecolor{darkyellow}{rgb}{0.5, 0.5, 0.0}
\newcommand{\sen}[1]{\textcolor{black}{#1}}
\newcommand{\andre}[1]{\textcolor{black}{#1}}
\newcommand{\newnew}[1]{\textcolor{black}{#1}}

\usepackage{xcolor,pifont}
\newcommand*\colourcheck[1]{%
  \expandafter\newcommand\csname #1check\endcsname{\textcolor{#1}{\ding{52}}}%
  \expandafter\newcommand\csname #1cross\endcsname{\textcolor{#1}{\ding{55}}}%
  \expandafter\newcommand\csname #1exclamation\endcsname{\textcolor{#1}{\textbf{!}}}%
}
\colourcheck{green}
\colourcheck{darkgreen}
\colourcheck{red}
\colourcheck{darkred}
\colourcheck{yellow}
\colourcheck{darkyellow}
\colourcheck{darkgrey}
\colourcheck{grey}
\colourcheck{orange}

\makeatletter
\newcommand\footnoteref[1]{\protected@xdef\@thefnmark{\ref{#1}}\@footnotemark}
\makeatother

\begin{document}

\title{RepairLLaMA: Efficient Representations and Fine-Tuned Adapters for Program Repair}

\author{André~Silva$^*$, Sen~Fang$^*$, and~Martin~Monperrus
\IEEEcompsocitemizethanks{\IEEEcompsocthanksitem A. Silva, M. Monperrus are with KTH Royal Institute of Technology, Sweden. S. Fang is with NC State University, USA.}
\thanks{E-mail: \{andreans, monperrus\}@kth.se, sfang9@ncsu.edu}
\thanks{$^*$ These authors contributed equally to this work.}}

\maketitle

\begin{abstract}
Automated Program Repair (APR) has evolved significantly with the advent of Large Language Models (LLMs).
Fine-tuning LLMs for program repair is a recent avenue of research, with many dimensions which have not been explored.
Existing work mostly fine-tune LLMs with naive code representations and does not scale to frontier models. 
To address this problem, we propose RepairLLaMA, a novel program repair approach that  
1) identifies optimal code representations for APR with fine-tuned models, and 
2) pioneers state-of-the-art parameter-efficient fine-tuning technique (PEFT) for program repair.
This results in RepairLLaMA producing a highly effective `program repair adapter' for fixing bugs with AI. 
Our experiments demonstrate the validity of both concepts.
First, fine-tuning adapters with program repair specific code representations enables the model to use meaningful repair signals and produce better patches.
Second, parameter-efficient fine-tuning helps fine-tuning to converge and clearly contributes to the effectiveness of RepairLLaMA in fixing bugs outside the fine-tuning data distribution.
Overall, RepairLLaMA correctly fixes 144 Defects4J v2, 109 HumanEval-Java, \new{and 20 GitBug-Java bugs}, outperforming all baselines.
\end{abstract}


\maketitle

\section{Introduction}

Automated program repair (APR) \cite{monperrus2018automatic, le2021automatic} aims at automatically fixing a software bug without human intervention.
Learning-based repair \cite{chen2019sequencer, Recoder, CURE-icse21, ye2022selfapr, xia2022less, jiang2023impact, xia2023plastic, wang2023rap} has become the mainstream solution to this problem due to the powerful ability of deep neural networks to learn complex bug fix patterns.
\new{Large language models (LLMs), pre-trained on vast amounts of data, have pushed learning-based repair to the next frontier \cite{jiang2023impact, xia2022practical}.}
\new{There are two main research lines in applying LLMs to program repair: (1) fine-tuning techniques to adapt pre-trained LLMs and create specialized repair models \cite{CURE-icse21, yuan2022circle, jiang2023impact, wang2023rap}, and (2) prompt engineering and agent-based approaches that leverage LLMs' capabilities through carefully designed inputs and interaction patterns \cite{joshi2023repair, xia2023keep, yang2024swe, zhang2024autocoderover, bouzenia2024repairagent}.}

Fine-tuning has the potential to learn domain-specific representations and leverage signals in order to generate high-quality patches. The drawback is that it requires substantial computational resources and high-quality training data.
Prompt engineering and agent-based approaches mostly consist of ad hoc human-developed prompts and workflows to instruct models for specific tasks and are bounded by the limited size of context windows.
\sen{
While both approaches offer different trade-offs, we argue that fine-tuning is crucial for high-quality automated program repair because: (1) it allows the model to internalize repair patterns from a supervised corpus of bug fixes, (2) it enables learning of task-specific repair representations rather than relying on manual prompt engineering steps, and (3) it provides opportunities to incorporate repair-specific information during training. Our paper is a key contribution in the space of fine-tuning LLMs for program repair.
}

Fine-tuning LLMs for program repair is complex. Early work simply refines the network weights based on additional fine-tuning data. However, this kind of fine-tuning is rather primitive and suffers from two significant drawbacks.
First, fine-tuning is also known to be able to adapt the input/output representations of the data under study \cite{chen2021plur}. In the context of program repair, there is an opportunity to fine-tune with code representations that maximize downstream task performance, that is, repair performance.
In particular, previous work overlooks the realistic representation of fault localization in the input.
Second, previous work considered the most basic fine-tuning technique, which is full-parameter fine-tuning.
As LLMs increase in size \cite{lu2023llama}, full-parameter fine-tuning poses important overfitting problems when fine-tuning data is limited, which is typically the case in program repair.
In this paper, we address the problem of devising efficient fine-tuning techniques \cite{hu2021lora} for program repair, with a focus on code representations and adapters.

We propose RepairLLaMA, a new program repair approach that \new{leverages parameter-efficient fine-tuning to adapt LLMs to handle repair-specific code representations.}
First, RepairLLaMA's code representations incorporate fault localization signals and are designed to support multi-location bugs.
Second, RepairLLaMA utilizes Low-Rank Adaption (LoRA), a state-of-the-art parameter-efficient fine-tuning technique, to train a much smaller \emph{repair adapter} (when compared to the full LLM) that adapts the LLM for program repair while helping prevent overfitting \cite{fu2023effectiveness}.
As we will demonstrate in this paper, the concept of \emph{repair adapter} is novel and potent. 

\new{Our experimental results validate RepairLLaMA's core design.
First, RepairLLaMA achieves state-of-the-art fine-tuning performance in three benchmarks, correctly fixing 144 Defects4J v2 \cite{defects4j} bugs, and 109 and 20 bugs, respectively, on recently proposed benchmarks HumanEval-Java \cite{jiang2023impact} and GitBug-Java \cite{silva2024gitbug}}, which boosts internal and external validity.
The experiments show that the devised code representations with repair signals allow the LLM to synthesize patches more effectively than the naive code-only representations. 
Also, RepairLLaMA clearly outperforms non-fine-tuned baselines, incl. GPT-4.
Moreover, our results also show the effectiveness of parameter-efficient fine-tuning:
RepairLLaMA's repair adapters, with only 4M parameters, are 1600x smaller than the initial pre-trained LLM, CodeLLama-7B \cite{roziere2023code}.
To sum up, the efficient representations and repair adapters of RepairLLaMA outperform recent results on fine-tuning for program repair \cite{jiang2023impact, huang2023empirical, wang2023rap} as well as world-class models such as GPT-3.5 and GPT-4.

Overall, we make the following contributions:
\begin{itemize}
    \item We design RepairLLaMA, an original fine-tuning pipeline for automated program repair with LLMs. RepairLLaMA's representations maximize knowledge from the program repair domain, while keeping strong alignment with pre-training.
    \item We systematically evaluate different code representations for program repair fine-tuning. Our results clearly show that the best code representation leverages the task-specific signals of fault localization and original buggy code. 
    \item We demonstrate that parameter-efficient fine-tuning performs \andre{competitively with} full-parameter fine-tuning in the context of program repair. The ``repair adapters'' of RepairLLaMA are training-efficient, and achieve state-of-the-art repair performance across \new{three} benchmarks, Defects4J, HumanEval-Java, \new{and GitBug-Java}, outperforming even GPT-4.
    \item For the sake of open science, we publish our source code, models, and artifacts at \url{https://github.com/ASSERT-KTH/repairllama}.
\end{itemize}

\begin{figure*}[t!]
    \centering
    \includegraphics[width=0.8\textwidth]{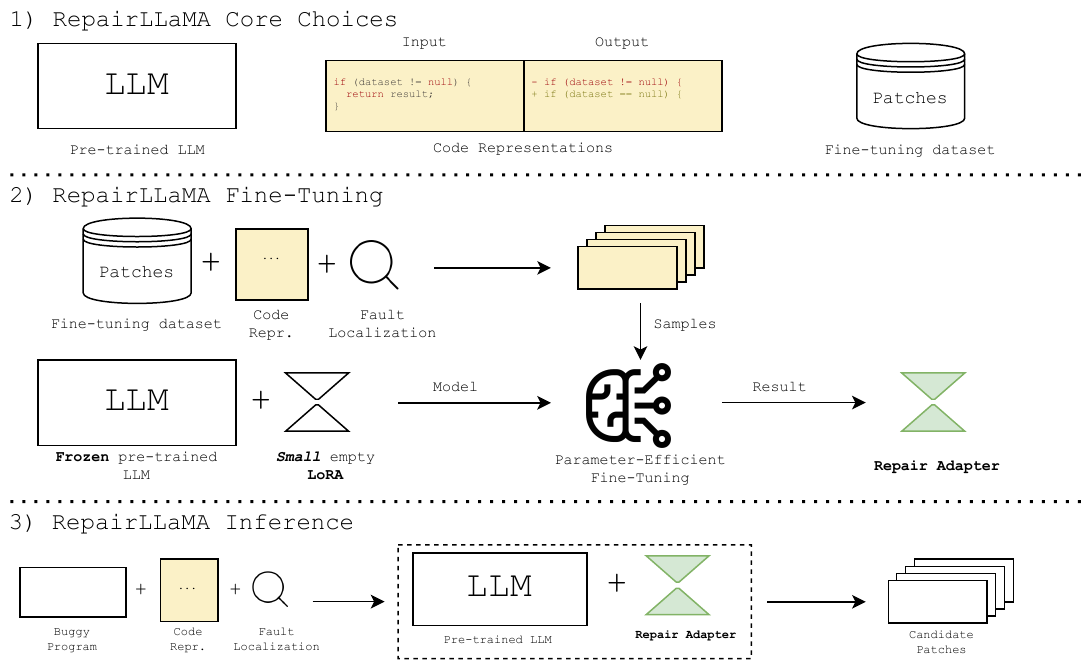}
    \caption{Overview of RepairLLaMA. The core novelties of RepairLLaMA are the APR specific code representations and the engineering of an effective \emph{program repair adapter} that is plugged into the underlying LLM.}
    \label{fig:overview}
\end{figure*}

\section{RepairLLaMA: Efficient Fine-Tuning for Program Repair}

\subsection{Overview}
\autoref{fig:overview} illustrates the pipeline of RepairLLaMA for APR, which is divided into three consecutive stages. The core novelties of this pipeline are:
1) the APR specific code representations, and 
2) the end-to-end use of a parameter-efficient fine-tuning technique.

The core of RepairLLaMA is a \emph{repair adapter}.
A \emph{repair adapter} is a plug-and-play extension of the model parameters that modifies the behavior of the LLM in order to maximize performance on the repair task, for a given programming language.
The adapter is responsible for transforming a rich, tailored input representation of the buggy code into the fit output representation of the patch. 

In the first stage of RepairLLaMA, the core choices are made, namely:
1) the initial pre-trained model (\autoref{sec:init-model});
2) the input code representation and output code representation (\autoref{sec:coderepr}); and
3) the fine-tuning dataset (\autoref{sec:finetuningdata}).
These choices are all important and are further discussed in the remainder of this section.

In the second stage, a repair adapter is trained.
The repair adapter is a much smaller (i.e., approx. 4M parameters) plug-and-play adapter of the initial LLM while remaining competitive on the task of program repair.

Finally, in the third stage, the repair adapter is employed to fix real-world bugs.

\subsection{Target Bugs}
\label{sec:target-bugs}

The first consideration when designing a fine-tuning pipeline for program repair is the bugs we aim to fix.
This relates to
1) the programming language,
2) the type of bugs (syntax errors, runtime errors, functional errors, etc), and
3) the difficulty of bugs, which can be proxied by the code span to modify in order to fix the bug.

In this work, we focus on 
1) Java bugs,
2) that are functional, and come with at least a failing test case,
3) that are intra-procedural, i.e. fixed with changes to a single function (called hereafter \textit{single-function bugs}). We do not make any assumption on the length of the method under repair, and
4) explicitly support bugs that require changes in multiple locations in the function \cite{ye2023iter}, beyond single-line or single-chunk bugs.

\subsection{Choice of the Initial LLM}
\label{sec:init-model}

Choosing the suitable initial LLM for fine-tuning is crucial. For example, when fine-tuning for code-related tasks, an LLM pre-trained on large-scale code corpora is more effective than one pre-trained on pure natural language data.
To effectively fine-tune an LLM for APR, we curate three criteria to choose the initial model.

First, the LLM should be publicly available and open-source. Fine-tuning a closed-source LLM on the task-specific dataset is not a valid option. Although some companies like OpenAI do provide an API for fine-tuning their LLMs, it is expensive, and the ownership of the final model (incl. weights) does not meet open-science reproduction criteria.
Open-source models, such as LLaMA~\cite{touvron2023llama} or StarCoder~\cite{li2023starcoder}, publish model weights online, allowing anyone to modify and deploy them.

Second, the LLM should be pre-trained with large-scale code data. As observed by related work \cite{li2023starcoder, roziere2023code}, LLMs pre-trained on massive code data achieve better performance in code-related tasks. Thus, we consider only LLMs specialized on code.

Third, the initial LLM should have been trained with an \infilling objective \cite{bavarian2022efficient} during pre-training.
As observed by related work \cite{jiang2023impact}, \infilling is a natural and effective learning objective for the program repair task, since it allows the model to synthesize code according to both the context appearing before and after.
It should also be supported by an off-the-shelf parameter-efficient fine-tuning library.

In \autoref{sec:exp:imp} we instantiate those criteria in the context of functional program repair for Java.

\subsection{Choice of Code Representations}
\label{sec:coderepr}

Source code representation is a critical aspect that significantly impacts the effectiveness of the model~\cite{namavar2022controlled}.
In this section, we discuss key characteristics of the source code representation design space.
We introduce, motivate, and elaborate on input and output code representations specific to the program repair task.

\subsubsection{Representation of Fault Localization}
\label{sec:fl}

Virtually all the APR literature assumes line-based fault localization, with a single line given as input to the repair algorithm.
This is not appropriate to fix multi-location bugs \cite{hercules, ye2023iter}.
Consider \autoref{fig:outputrepr} (OR4), which shows the canonical patch for the multi-location bug Chart-5 from Defects4J.
In this case, fault localization must identify a location where an entirely new if block should be synthesized and inserted as well as another pre-existing if condition, appearing later in the code.
To our knowledge, there is no fault localization technique able to predict tuples of blocks to be repaired together. 

In this paper, we propose a novel way to represent fault localization information: our core idea is to represent fault localization not as a single line, but as a region.
In RepairLLaMA, we encode fault localization as a span ranging from the beginning of the suspicious region to its end.
This encoding is realistic because 1) identifying a buggy method is within reach of existing fault localization methods, and 2) exhaustively listing all suspicious code regions of a buggy method is worst-case $O(n^2)$ in the number of method lines.

\subsubsection{Input Representation Space}
\label{sec:inputrepr}

\begin{figure}[t]
    \centering
    \includegraphics[width=0.425\textwidth]{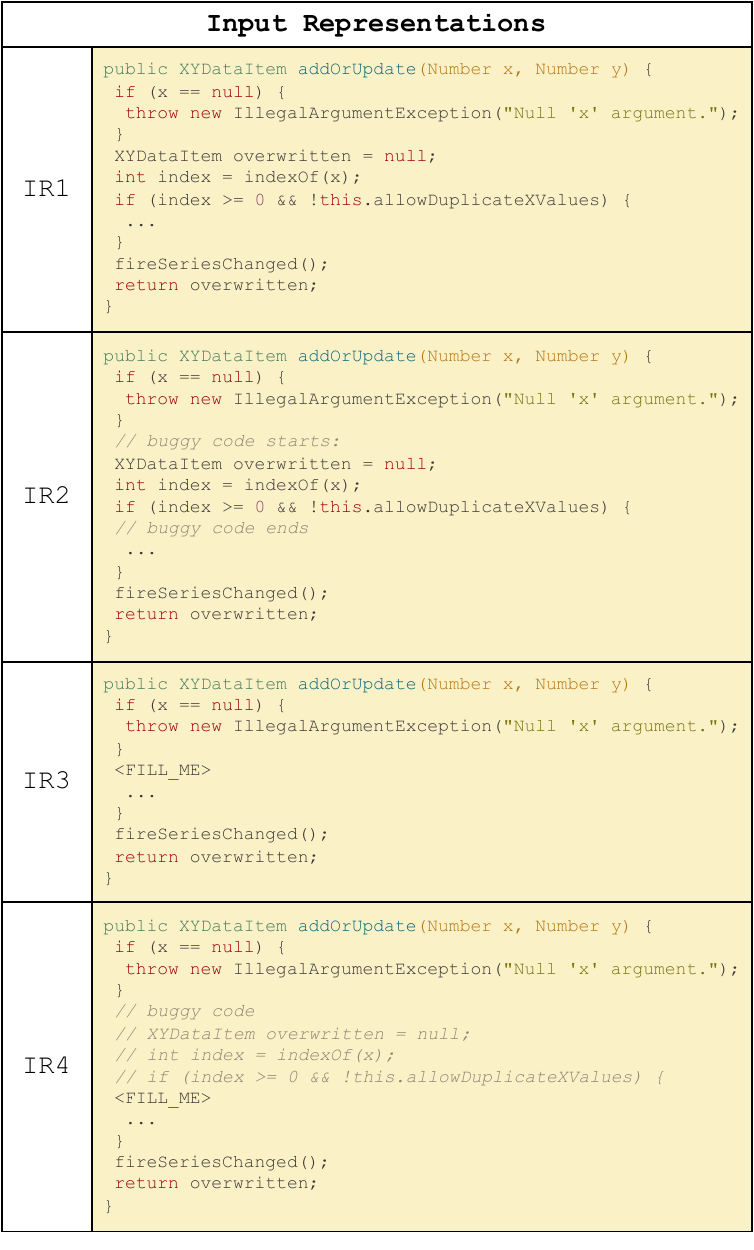}
    \caption{Buggy code of the multi-location bug Chart-5 represented in our four different input representations.}
    \label{fig:inputrepr}
\end{figure}

In APR, the design space of the input representation relates to what is shown from the buggy code and to the presence of additional information.
For example, fault localization signals can be useful in scoping down where the code should be modified.
However, such information might not be seen at the pre-training stage.
For the LLM to utilize it, one must represent it in a way that it can learn during fine-tuning.
To study the input representation space, we design four input representations tailored to APR (\autoref{fig:inputrepr}):

\textbf{IR1: Buggy function}
This naive representation describes the code in the standard format as it is written, simply as text.
\autoref{fig:inputrepr} (IR1) shows the buggy function of the multi-location bug Chart-5, a Defects4J bug.
The advantage of IR1 is that it is the same representation LLMs observe during pre-training.
When using this representation, the main limitation is that the model has no access to fault localization information and, thus, needs to determine where to change the code, which can be considered as implicit anomaly detection.

\textbf{IR2: Buggy function w/ FL comments}
This representation adds two comments signaling the start and end of the buggy chunk of code.
For example, in \autoref{fig:inputrepr} (IR2), the three lines between the start and end of the suspicious region are surrounded by comments signaling the beginning and end of the buggy chunk.
By providing fault localization information, the model can scope its changes to the buggy section.

\textbf{IR3: Buggy function w/ \infilling mask}
This representation uses the \infilling scheme some LLMs are trained for during pre-training~\cite{bavarian2022efficient}. 
The buggy chunk is replaced by the infilling token, which prompts the model to fill it.
For example, in \autoref{fig:inputrepr} (IR3), the three lines between the start and end of the suspicious region are replaced by the \textit{<FILL\_ME>} token.
This representation yields shorter inputs and requires less fine-tuning since the \infilling objective has been used during pre-training.
However, by masking the buggy portion of code, this representation incurs information loss that can be useful to generate a fix.

\textbf{IR4: Buggy function w/ \infilling mask and buggy code}
This representation combines the buggy code with the \infilling scheme.
The buggy code is shown in a comment at the end of the prefix portion.
For example, in \autoref{fig:inputrepr} (IR4), the buggy lines are kept in comments, and the \texttt{<FILL\_ME>} token is placed immediately afterward.
This representation is different from the one learned during pre-training and requires fine-tuning.
Code found in the wild would typically not include buggy code as comments, which is considered bad practice.
Yet, with fine-tuning, this representation might add valuable information to the \infilling scheme.

\subsubsection{Output Representation Space}
\label{sec:outputrepr}

\begin{figure}[t]
    \centering
    \includegraphics[width=0.425\textwidth]{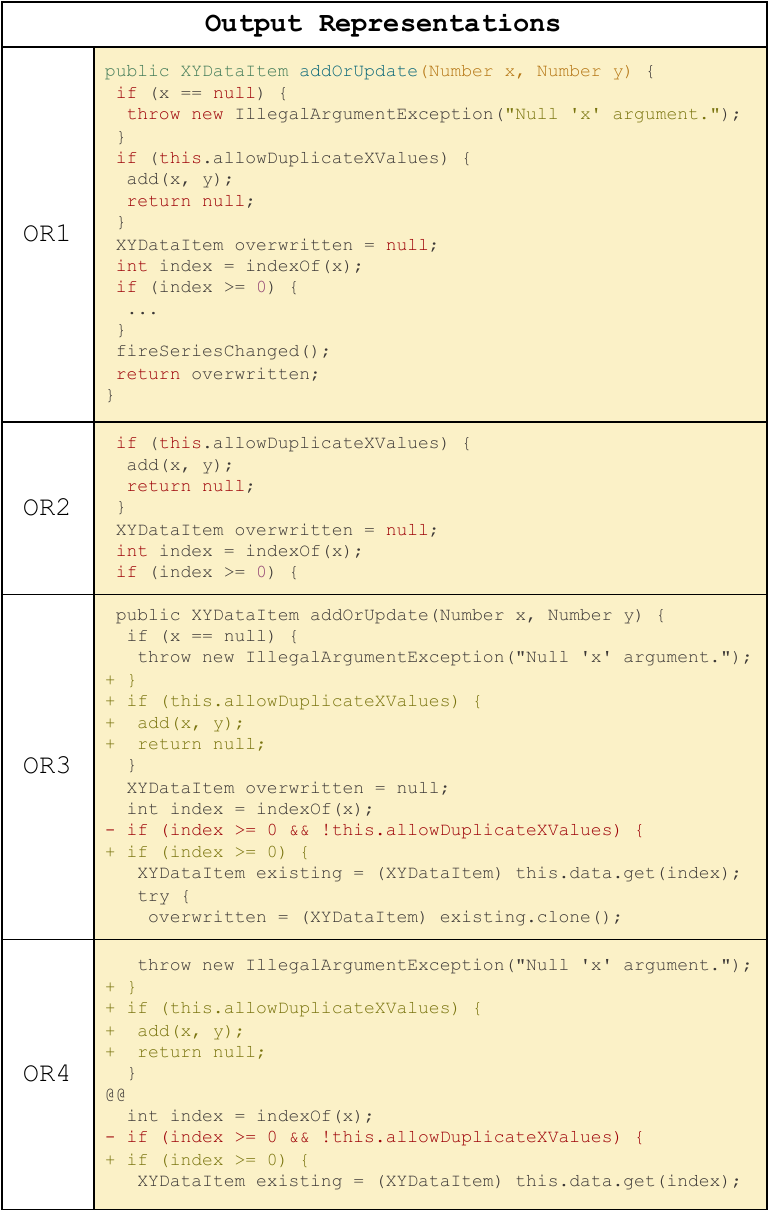}
    \caption{Patch for multi-location bug Chart-5 represented in our four different output representations.}
    \label{fig:outputrepr}
\end{figure}

Output representations in APR correspond to the representation of the synthesized fixed code.
A natural output representation is a diff over the buggy code, aka a patch.
As discussed in \autoref{sec:inputrepr}, fine-tuning is required to adapt an LLM to generate such task-specific outputs.
To study the output representation space, we design four output representations tailored to APR (\autoref{fig:outputrepr}):

\textbf{OR1: Fixed function}
The naive output is the full fixed function.
It is not a diff.
\autoref{fig:outputrepr} (OR1) shows the fixed function of the multi-location bug Chart-5.
The major drawback of OR1 is that such output may be much larger than the actual code changes for fixing, and LLMs are known to be more effective at generating short sequences over long sequences.

\textbf{OR2: Fixed chunk}
In this representation, the output is composed of the fixed chunk of code to replace the buggy chunk of code. 
The advantage is that the fixed chunk is typically shorter than the full function, i.e. shorter than OR1.
For example, in \autoref{fig:outputrepr} (OR2), only 6 fixed lines are outputted.
OR2 requires an input representation that includes fault localization (i.e. IR2, IR3, IR4) since the output contains no information regarding what to replace.

\textbf{OR3: Three-line context-diff}
The output is a typical contextual diff with a three-line context, aka a unified diff.
For example, in \autoref{fig:outputrepr} (OR3), a unified diff of the statement change is outputted.
The main challenge of this representation is that the model needs to learn to locate the bug locations during fine-tuning, which is difficult.
Additionally, this representation is also lengthier than generating a fixed chunk (OR2) only.

\textbf{OR4: One-line context-diff}
The output is a contextual diff with a shorter, one-line context.
OR4 uses a one-line diff context, making it shorter than OR3.
For example, in \autoref{fig:outputrepr} (OR4), there are five source code lines less when compared with OR3.
Despite this, it is still lengthier than OR2 and also requires the model to learn where to apply the patch.

\subsubsection{Input/Output Representation Pairs}

\begin{table}[t]
\centering
\caption{Possible code representation pairs for fine-tuning LLMs for automated program repair. They exploit the characteristics of the APR task, incl. the presence of fault localization signals and the notion of ``buggy code''.}
\label{tab:code_representations}
\resizebox{0.9\linewidth}{!}{%
\begin{tabular}{|c|c|c|c|} 
\hline
\textbf{Code Representations} & \textbf{FL} & \textbf{Aligned w/ PT} & \textbf{Buggy Code}  \\ 
\hline
IR1 x OR1                     & \redcross   & \greencheck / \redcross           & \greencheck          \\ 
\hline
IR1 x OR3                     & \redcross   & \greencheck / \redcross              & \greencheck          \\ 
\hline
IR1 x OR4                     & \redcross   & \greencheck / \redcross              & \greencheck          \\ 
\hline
IR2 x OR2                     & \greencheck & \redcross / \greencheck              & \greencheck          \\ 
\hline
IR3 x OR2                     & \greencheck & \greencheck / \greencheck            & \redcross            \\ 
\hline
IR4 x OR2                     & \greencheck & \redcross / \greencheck           & \greencheck          \\
\hline
\end{tabular}
}
\end{table}

To utilize an LLM for APR, input and output representations must be carefully paired. This is because all input representations cannot be paired with all output representations. For instance, IR1 cannot pair with OR2 since one cannot apply a fixed chunk to the buggy function without the fault localization information.
\autoref{tab:code_representations} provides the list of the code representation pairs that are studied in this paper.
Each row corresponds to a code representation pair.
Column \textit{FL} indicates whether the pair includes or not fault localization information.
Column \textit{Aligned w/ PT} provides a relative assessment of the alignment of the representation w.r.t. the pre-training data/objective. A red cross means that the code representation is not aligned with the pre-training data and objective. The left side shows the input and the right the output representations.
Column \textit{Buggy Code} indicates whether the pair includes or not the original buggy code.

The first three rows (i.e., IR1xOR1, IR1xOR3, IR1xOR4) include code representation pairs that do not contain fault localization signals.
The input is the same across all pairs (IR1), whereas the output can either be the full fixed function (OR1) or a diff (OR3, OR4).
The key difference between the pairs is the output length and format.

The latter three rows (i.e., IR2xOR2, IR3xOR2, IR4xOR2) include code representation pairs that contain fault localization information, either as tokens or as infilling, which is specific to program repair.
The most aligned representation with pre-training is IR3xOR2 since the pre-trained model has support for \infilling.
IR2 represents the \infilling objective with never-before-seen comments, whereas IR4 keeps the buggy code as comments. 
The natural output representation to pair with these is OR2 since it only includes the new code to replace the already localized buggy chunk, minimizing output length.
Note that we have empirically tested other combinations in a pilot experiment, and the ones not listed in \autoref{tab:code_representations} underperform.

\subsection{Choice of Fine-Tuning Dataset}
\label{sec:finetuningdata}

After choosing an initial model and appropriate code representations, the next step is to curate a fine-tuning dataset.
First, the dataset must be relevant to the task at hand.
In the APR task, a relevant dataset usually includes pairs of buggy and fixed code samples.
Second, the type of samples included should be similar to the target bugs.
Third, the size of the dataset should be considered.
A larger dataset generally leads to better model performance as it provides more examples for the model to fine-tune from.
However, it is important to balance size with quality - a smaller, high-quality dataset may be more beneficial than a larger, low-quality one.
Fourth, the diversity of the dataset is important.
A diverse dataset that covers a wide range of examples can help the model generalize better to unseen data.
Lastly, the legality and ethics of the dataset should be considered, in particular regarding privacy and copyright.

\subsection{Program Repair Adapters for LLMs}

With the recent release of various LLMs, the scale of parameters has significantly increased. For instance, state-of-the-art models such as LLaMA \cite{lu2023llama} and CodeLLaMA \cite{roziere2023code} range from 7B to 70B parameters. Fine-tuning these LLMs often requires substantial GPU resources. As an example, \citeauthor{lv2023full} \cite{lv2023full} report that fine-tuning the full parameters of LLaMA-7B on an RTX 3090 consumes 126.08 GB at peak GPU memory usage, with the batch size and sequence length set to 1 and 1024 respectively. Fine-tuning current LLMs with limited resources is a challenge.

\new{RepairLLaMA uses LoRA \cite{hu2021lora}, a state-of-the-art parameter-efficient fine-tuning method that reduces memory requirements while maintaining model performance. Instead of fine-tuning all model parameters, LoRA freezes the pre-trained LLM weights and injects trainable low-rank matrices into specific attention layers.
In addition to reducing memory requirements, by reducing the number of trainable parameters LoRA also acts as a regularizer \cite{fu2023effectiveness} helping prevent overfitting during fine-tuning.}

\new{In our implementation, we apply LoRA to the query (\texttt{q\_proj}) and value (\texttt{v\_proj}) projection matrices in each transformer's self-attention layer. These projection matrices transform the input embeddings into query and value vectors used in the attention. For each target weight matrix $A \in \mathbb{R}^{d \times k}$, LoRA decomposes the update into a product of two lower-rank matrices $B \in \mathbb{R}^{d \times r}$ and $C \in \mathbb{R}^{r \times k}$, where $r$ is the rank. During inference, the effective weight matrix is $W = W_0 + BA$, where $W_0$ is the frozen pre-trained weight.}

\new{The trained matrices can be interpreted as a \emph{repair adapter} that is much smaller than the original LLMs. For example, with rank r=8, our adapter only requires training 0.39\%\footnote{Take CodeLLaMA-7B as an example, its hidden size is 4096, so the original parameter of a matrix is 4096 $\times$ 4096. As for the LoRA adapter, this value is 4096 $\times$ 8 + 4096 $\times$ 8.} of the parameters compared to full fine-tuning. This parameter-efficient approach allows RepairLLaMA to achieve strong program repair performance while being trainable on a single GPU.}

\subsection{Inference Time}

The final step is to deploy the repair adapter. The target buggy program is fed to a fault localization algorithm and processed to generate an APR-specific code representation.
Then, the code representation is fed to the initial model combined with the LoRA repair adapter to generate a list of candidate patches for the buggy program. Patches are then checked for plausibility and correctness per off-the-shelf techniques.

\section{Experimental Methodology}

\subsection{Research Questions} In this work, we focus on the following research questions:
\begin{itemize}[wide, labelwidth=!, labelindent=0pt]
    \item \textbf{RQ1 (Code Representations for Fine-Tuning)}:
    What is the best code representation to fine-tune an LLM for program repair?

    \item \textbf{RQ2 (Parameter-Efficient Fine-Tuning vs. Full Fine-Tuning)}:
    How does parameter-efficient fine-tuning compare against full-parameter fine-tuning for program repair?

    \item \textbf{RQ3 (RepairLLaMA vs. ChatGPT-based APR)}:
    How does RepairLLaMA compare against state-of-the-art ChatGPT-based program repair?
\end{itemize}

\subsection{Implementation}
\label{sec:exp:imp}

\noindent\textbf{Model to Fine-Tune}
Per the criteria of \autoref{sec:init-model}, we choose CodeLlama-7b \cite{roziere2023code} as our initial LLM.
CodeLLaMA is a publicly available LLM released in 2023 and is trained on 500B code tokens. 
Per the experiments reported in \cite{roziere2023code}, CodeLLaMA outperforms GPT-3.5 on two code generation benchmarks.

\noindent\textbf{Fine-tuning Dataset}
We choose Megadiff \cite{monperrus2021megadiff} as the fine-tuning dataset, and process all samples into the different code representations.
First, the function pairs -- each comprising a buggy version and its fixed counterpart -- are extracted along with their corresponding diff identifiers.
Subsequently, we eliminate pairs that do not change single functions, and remove duplicate pairs through textual comparison.
After that, we compute our custom code representations.
We keep only samples whose total length (input plus output) is shorter than 1024 tokens measured by the LLM tokenizer. 
Consequently, the fine-tuning datasets range from 30,000 to 50,000 fine-tuning pairs (see our appendix repository).

\noindent\textbf{Evaluation Benchmark}
We select \new{three} Java benchmarks for our evaluation: Defects4J \cite{defects4j}, HumanEval-Java \cite{jiang2023impact}, and GitBug-Java \cite{silva2024gitbug}.
Following recent related work~\cite {jiang2021cure, xia2022practical, xia2023keep}, we scope our evaluation to single-function bugs, as defined in \autoref{sec:target-bugs}.
Defects4J comprises 835 real-world bugs from 17 open-source Java projects, from which we identify 488 single-function bugs.
HumanEval-Java is a bug benchmark containing artificial bugs inserted in HumanEval~\cite{chen2021evaluating} Java programs.
HumanEval-Java contains 162 single-function bugs.
\new{GitBug-Java is a bug benchmark of recent bugs, collected from the 2023 commit history of 55 open-source repositories, comprising 90 single-function bugs.}
Contrary to Defects4J, \new{GitBug-Java} suffers from less data leakage in the pre-training data since \new{all bugs are} much more recent than Defects4J.
\new{Notably, GitBug-Java exclusively contains bugs from after the training data cutoff date of all models used in our experiments: CodeLlama-7b (September 2022) \cite{roziere2023code}, \textit{gpt-4-0613} (September 2021) \footnote{\label{openai-cutoff-note}\url{https://platform.openai.com/docs/models}}, and \textit{gpt-3.5-turbo-0613} (September 2021) \footnoteref{openai-cutoff-note}.}

\noindent\textbf{Fine-Tuning Hyperparameters}
We fine-tune CodeLLaMA with LoRA for each of our curated code representations with the same hyper-parameter settings: we set the learning rate to 5e-4 with cosine decay, max input length to 1024\footnote{\new{Although longer input lengths enable the model to process more context and, potentially, fix more bugs, about four times more GPU memory is required when doubling the input length.}}, training epoch to 2, and batch size to 16 per GPU, and we use Adam\_W as the optimizer.
For LoRA, we use a rank of 8, alpha of 16, dropout of 0.05, and inject the adaptation matrices in the \textit{q\_proj} and \textit{v\_proj} layers.
Using the same hyper-parameter settings for each code representation ensures fair comparison.
Each fine-tuning run is executed on a server with 4xA100 40GB GPUs.

\noindent\textbf{Inference Setup}
In inference, we employ beam search as our decoding strategy with a beam size of 10 per previous research \cite{jiang2023impact}.
Hence, for each bug, we generate 10 candidate patches.
We use the HuggingFace transformers library to implement all fine-tuning and inference experiments.
Inference is run on a single A100 40GB GPU.

\subsection{Patch Assessment}
\label{sec:exp:assess}

Patch assessment is a notoriously hard task.
In our evaluation, we compute the best metrics for that task \cite{xia2022practical, jiang2023impact, ye2023iter}:
1) A \textit{plausible patch} is defined as one that successfully passes all test cases.
2) An \textit{exact-match patch} is textually identical to a developer-provided reference patch, incl. spaces and formatting.
3) An \textit{AST-match patch} has an AST which is equivalent to the AST of the developer-provided reference patch.
4) A \textit{semantic-match patch} is a patch deemed equivalent after manual assessment by an expert.

Plausible and exact-match patches are straightforward. Let us dwell on the two other kinds.

The major advantage of an \textit{AST-match} patch is to compute performance regardless of formatting and indentation changes. 
It is also more scalable than manually checking patches for correctness without expertise in the programs under repair.
The \textit{AST-match} process involves converting plausible and reference patches into abstract syntax trees \cite{spoon} and subsequently utilizing AST differencing \cite{falleri2014fine} to compare their ASTs for discrepancies.  

A \textit{semantic-match} patch is the most costly assessment to get. In this paper, to assess semantic equivalence, the two first authors independently label all plausible but not AST/Exact match patches in a first round. For the patches the two first authors disagree upon, the third author breaks the tie.

For all four metrics, the higher the metric, the better the performance.
We validate the candidate patches on a workstation with an 18-core Intel Core i9-10980XE CPU and 128 GB of RAM, operating under Ubuntu 22.04.3 LTS.

\subsection{Methodology for RQ1}

\label{sec:exp:rq1}

The objective of RQ1 is to investigate the most effective code representations for fine-tuning an LLM for program repair.
While existing research has delved into the utility of LLMs for program repair, the impact of the code representations, such as their realism, has been overlooked.
It is known that variations in code representations may yield substantial differences in performance for fine-tuned LLMs \cite{huang2023empirical}.
Consequently, in RQ1, we empirically evaluate 6 realistic code representation pairs presented in \ref{sec:coderepr} and measure their performance.

We fine-tune an LLM as described in \ref{sec:exp:imp}.
We prompt the model to generate 10 patches for each bug using beam search decoding.
We then evaluate the generated patches as outlined in \autoref{sec:exp:assess}, to measure the effectiveness of each code representation.
We prompt the non fine-tuned CodeLLaMA-7B as a baseline.

\new{We assess the statistical significance of performance difference by employing the McNemar test for each pairwise combination of representations.
In our context, each statistical test looks at the binary outcomes of two representations (or models) evaluated on the same set of benchmark examples, according to semantical match.
The null hypothesis ($H_0$) for the McNemar's tests is that the distributions are indistinguishable, i.e. the row and column marginal frequencies in the 2x2 contingency table are equal, i.e., $H_0: P(A=1, B=0) = P(A=0, B=1)$, where A and B represent whether A and B, respectively, fixed a given bug.
}

\subsection{Methodology for RQ2}
\label{sec:exp:rq2}

The objective of RQ2 is to evaluate the respective effectiveness of parameter-efficient and full-parameter fine-tuning.
Generally, parameter-efficient fine-tuning methods represent a trade-off between computational cost and model performance, in order to train LLMs with limited computational resources.
While traditional full-parameter fine-tuning approaches often yield better results, it comes at the expense of significantly higher memory requirements and a larger-scale fine-tuning dataset. In other words, fully fine-tuning an LLM on a small fine-tuning dataset may result in overfitting.
In this experiment, we explore and compare the effectiveness of parameter-efficient and full-parameter fine-tuning in the specific context of program repair, which has never been done to the best of our knowledge.

\emph{Baseline.}
\andre{
We consider two baseline base models in RQ2: CodeLLaMA-7B  and deepseek-coder-6.7b-base~\cite{guo2024deepseek}.
To study the difference between parameter-efficient fine-tuning and full-parameter fine-tuning, we fine-tune both models with both approaches.
Here, we use the same hyper-parameters as in LoRA fine-tuning described in RQ1. We use a learning rate of 2e-5 for full-parameter-fine-tuning.
We employ the same statistical significance test as described in \autoref{sec:exp:rq1} when comparing RepairLlama against these baselines.
}

\sen{Additionally, we benchmark our approach against the highest-performing model documented by Jiang et al. \cite{jiang2021cure}—specifically, the fine-tuned variant of Incoder-6B \cite{fried2022incoder}. We also compare with RAP-Gen \cite{wang2023rap}, a state-of-the-art model for automated program repair that enhances processing of buggy inputs by leveraging fix patterns retrieved from an extensive repository of historical bug-fix pairs.}

\subsection{Methodology for RQ3}
\label{sec:exp:rq3}

\begin{figure}[t]
    \centering
    \includegraphics[width=\columnwidth]{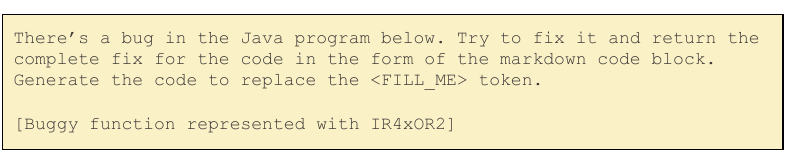}
    \caption{The prompt used to prompt GPT-3.5 and GPT-4 as a strong baseline to generate patches.}
    \label{fig:chatgpt_prompt}
\end{figure}

Recently, related work \cite{xia2023keep, zhang2023critical} has shown that GPT-3.5-Turbo and GPT-4 achieve state-of-the-art results on program repair.
The objective of RQ3 is to study how RepairLLaMA compares against state-of-the-art ChatGPT-based program repair. 

First, we zero-shot prompt \textit{gpt-3.5-turbo-0613} and \textit{gpt-4-0613} to generate 10 patches for each bug.
The prompt is shown in \autoref{fig:chatgpt_prompt}, \new{it is built by integrating the best code representation with an effective zero-shot prompt from \citeauthor{zhang2023critical}~\cite{zhang2023critical}. The prompt instructs the LLM} to generate the fixed code chunk to replace the $\text{<FILL\_ME>}$ token.
\new{Note that the same information signals are provided to both RepairLLaMA models and the GPT models, allowing for a fair comparison of the repair capabilities of the different models. This includes providing the same source code context, as well as the same fault localization information as input.}
\new{We employ the same statistical significance test as described in \autoref{sec:exp:rq1}.}
We utilize OpenAI's official APIs to call \textit{gpt-3.5-turbo-0613} and \textit{gpt-4-0613} to conduct our experiments on Dec. 1, 2023, \new{except for the GitBug-Java experiments which are conducted on \textit{gpt-3.5-turbo-0125} on Nov. 17, 2024}.

\section{Experimental Results}

\subsection{Results of RQ1 (Code Representations for Fine-Tuning)}

\begin{table*}
\centering
\caption{Repair results of different code representations for fine-tuning an LLM for program repair (RQ1). Our best model, RepairLLaMA using IR4xOR2, significantly improves over the baseline in \new{all} test benchmarks.}
\label{tab:rq1_results}
\resizebox{\linewidth}{!}{%
\begin{tblr}{
  row{2} = {c},
  column{2} = {c},
  cell{1}{1} = {r=2}{},
  cell{1}{2} = {r=2}{},
  cell{1}{3} = {c=4}{c},
  cell{1}{7} = {c=4}{c},
  cell{1}{11} = {c=4}{c},
  cell{3}{3} = {r},
  cell{3}{4} = {r},
  cell{3}{5} = {r},
  cell{3}{6} = {r},
  cell{3}{7} = {r},
  cell{3}{8} = {r},
  cell{3}{9} = {r},
  cell{3}{10} = {r},
  cell{3}{11} = {r},
  cell{3}{12} = {r},
  cell{3}{13} = {r},
  cell{3}{14} = {r},
  cell{4}{3} = {r},
  cell{4}{4} = {r},
  cell{4}{5} = {r},
  cell{4}{6} = {r},
  cell{4}{7} = {r},
  cell{4}{8} = {r},
  cell{4}{9} = {r},
  cell{4}{10} = {r},
  cell{4}{11} = {r},
  cell{4}{12} = {r},
  cell{4}{13} = {r},
  cell{4}{14} = {r},
  cell{5}{3} = {r},
  cell{5}{4} = {r},
  cell{5}{5} = {r},
  cell{5}{6} = {r},
  cell{5}{7} = {r},
  cell{5}{8} = {r},
  cell{5}{9} = {r},
  cell{5}{10} = {r},
  cell{5}{11} = {r},
  cell{5}{12} = {r},
  cell{5}{13} = {r},
  cell{5}{14} = {r},
  cell{6}{3} = {r},
  cell{6}{4} = {r},
  cell{6}{5} = {r},
  cell{6}{6} = {r},
  cell{6}{7} = {r},
  cell{6}{8} = {r},
  cell{6}{9} = {r},
  cell{6}{10} = {r},
  cell{6}{11} = {r},
  cell{6}{12} = {r},
  cell{6}{13} = {r},
  cell{6}{14} = {r},
  cell{7}{3} = {r},
  cell{7}{4} = {r},
  cell{7}{5} = {r},
  cell{7}{6} = {r},
  cell{7}{7} = {r},
  cell{7}{8} = {r},
  cell{7}{9} = {r},
  cell{7}{10} = {r},
  cell{7}{11} = {r},
  cell{7}{12} = {r},
  cell{7}{13} = {r},
  cell{7}{14} = {r},
  cell{8}{3} = {r},
  cell{8}{4} = {r},
  cell{8}{5} = {r},
  cell{8}{6} = {r},
  cell{8}{7} = {r},
  cell{8}{8} = {r},
  cell{8}{9} = {r},
  cell{8}{10} = {r},
  cell{8}{11} = {r},
  cell{8}{12} = {r},
  cell{8}{13} = {r},
  cell{8}{14} = {r},
  cell{9}{3} = {r},
  cell{9}{4} = {r},
  cell{9}{5} = {r},
  cell{9}{6} = {r},
  cell{9}{7} = {r},
  cell{9}{8} = {r},
  cell{9}{9} = {r},
  cell{9}{10} = {r},
  cell{9}{11} = {r},
  cell{9}{12} = {r},
  cell{9}{13} = {r},
  cell{9}{14} = {r},
  cell{10}{3} = {r},
  cell{10}{4} = {r},
  cell{10}{5} = {r},
  cell{10}{6} = {r},
  cell{10}{7} = {r},
  cell{10}{8} = {r},
  cell{10}{9} = {r},
  cell{10}{10} = {r},
  cell{10}{11} = {r},
  cell{10}{12} = {r},
  cell{10}{13} = {r},
  cell{10}{14} = {r},
  vline{2-15} = {1}{},
  vline{4-15} = {2}{},
  vline{2-15} = {2-10}{},
  hline{1,3-11} = {2-14}{},
  hline{2} = {3-14}{},
}
  & \textbf{Code Representations }   & {\textbf{Defects4J v2}\\\textbf{(488 bugs)}} &                &              &                   & {\textbf{HumanEval-Java}\\\textbf{(162 bugs)}} &                &              &                   & {\textbf{GitBug-Java}\\\textbf{(90 bugs)}} &                &              &                   \\
  &                                  & Plausible                                    & {Exact\\Match} & {AST\\Match} & {Semantic\\Match} & Plausible                                      & {Exact\\Match} & {AST\\Match} & {Semantic\\Match} & Plausible                                  & {Exact\\Match} & {AST\\Match} & {Semantic\\Match} \\
1 & IR3 x OR2 (no fine-tuning)       & 131                                          & 52             & 70           & 83                & 107                                            & 71             & 81           & 103               & 17                                         & 8              & 9            & 12                \\
2 & IR4 x OR2 (no fine-tuning)       & 107                                          & 50             & 60           & 69                & 95                                             & 65             & 72           & 91                & 19                                         & 11             & 12           & 13                \\
3 & IR1 x OR1                        & 79                                           & 29             & 31           & 45                & 78                                             & 52             & 54           & 72                & 10                                         & 4              & 4            & 4                 \\
4 & IR1 x OR3                        & 41                                           & 15             & 17           & 24                & 39                                             & 21             & 21           & 37                & 6                                          & 1              & 1            & 1                 \\
5 & IR1 x OR4                        & 12                                           & 2              & 2            & 3                 & 5                                              & 2              & 2            & 4                 & 1                                          & 1              & 1            & 1                 \\
6 & IR2 x OR2                        & \textbf{198}                                 & 121            & 122          & 139               & \textbf{118}                                   & 69             & 77           & 108               & 23                                         & \textbf{16}    & \textbf{17}  & 19                \\
7 & IR3 x OR2                        & 153                                          & 83             & 86           & 102               & 103                                            & 63             & 68           & 99                & 21                                         & 12             & 12           & 13                \\
8 & \textbf{IR4 x OR2 (RepairLLaMA)} & 195                                          & \textbf{124}   & \textbf{125} & \textbf{144}      & \textbf{118}                                   & \textbf{75}    & \textbf{82}  & \textbf{109}      & \textbf{25}                                & 15    & 16           & \textbf{20}                
\end{tblr}
}
\end{table*}

In RQ1, we investigate the most effective code representations for fine-tuning an LLM for program repair.
The results of the evaluation are presented in \autoref{tab:rq1_results}, which shows the effectiveness of each code representation setting on both test benchmarks.
The table is structured as follows:
the first column displays the code representations,
the second \new{to fourth} meta-columns show the repair effectiveness results on, respectively, \new{the single-function subsets of} Defects4J v2, HumanEval-Java, \new{and GitBug-Java}.
Recall that the repair effectiveness evaluation is measured by the four patch assessment metrics described in \autoref{sec:exp:assess}.

Our baselines (rows 1 and 2) use the original non-fine-tuned CodeLLaMA-7B, with the IR3xOR2 and IR4xOR2 code representations respectively.
We computed the baseline performance for other representations, and these are the most effective way to prompt the non-fine-tuned model \cite{jiang2023impact}, making them the strongest possible baselines.

Our results show that the IR3xOR2 baseline plausibly repairs 131 Defects4J v2, 107 HumanEval-Java bugs, \new{and 17 GitBug-Java bugs}.
Moreover, it correctly repairs 52 Defects4J v2, 71 HumanEval-Java, \new{and 8 GitBug-Java} bugs with patches that textually exactly match the developer-written ones.
Furthermore, when considering AST match, it can repair 70 Defects4J v2, 81 HumanEval-Java, \new{and 9 GitBug-Java} bugs with patches that are syntactically equal to the developer-written ones.
Finally, when considering semantic match, the baseline can repair 83 Defects4J v2, 103 HumanEval-Java, \new{and 12 GitBug-Java} bugs with patches that are semantically equivalent to the developer-written ones.
Those results show that CodeLLaMA-7B with the appropriate prompting is already a very strong model for program repair.

The difference between the first baseline (row 1, IR3xOR2) and the second baseline (row 2, IR4xOR2) is that IR4 includes the buggy code as a comment.
Despite this extra signal, the baseline model cannot make use of it effectively.
Indeed, it achieves worse performance, correctly repairing 14 (69 vs. 83 of IR3xOR2) less Defects4J bugs, 12 (91 vs. 103) less HumanEval-Java bugs, \new{and 1 (13 vs. 12) more GitBug-Java bug}.
This finding clearly shows that without any fine-tuning, inserting extra information into the representation such as the buggy code can act as noise.

The fine-tuned model's effectiveness depends on the code representations.
First, we observe that the three code representation pairs that do not have access to fault localization (IR1xORX, rows 3 to 5) perform considerably worse than both the baseline and other code representations.
These results show that fault localization signals are crucial for program repair. To that extent, all representations that simply use the full function as the input and ask to ``fix the bug'', can be considered too naive. 
Our work demonstrates that tailoring code representations with fault localization is a necessary step in the context of program repair.
Recall that no pre-training objective has access to fault-localization signals, which means that fine-tuning is essential to build knowledge of fault-localization signals.

Second, we observe that, within code representations that use fault localization signals (IR2, IR3, and IR4, rows 6 to 8), fine-tuned models significantly outperform the baseline on Defects4J v2 \new{and GitBug-Java} compared to the baselines (rows 1 and 2).
Again, this clearly demonstrates the importance of fault localization signals in the input representation.

\new{The only exception to this is in HumanEval-Java, where the fine-tuned IR3xOR2 model performs worse than the baseline IR3xOR2 model.
We hypothesize that this phenomenon occurs due to catastrophic forgetting: fine-tuning improves the model for a downstream task at the expense of some loss of knowledge from previous training.
In our case, the catastrophic forgetting of HumanEval-Java is explained by the difference in nature between the fine-tuning dataset and the benchmark.
Megadiff, the fine-tuning dataset, contains diffs sampled from Java projects hosted on GitHub.
The nature of such projects is closer to the nature of Defects4J and GitBug-Java samples, as these are also collected from open-source Java repositories. HumanEval-Java, on the other hand, is manually engineered on top of simple coding tasks not commonly found on large open-source projects, and thus has a different distribution of bug and fix kinds.
These results highlight the need for careful curation of fine-tuning datasets and benchmarks.
Additionally, it also supports the effectiveness of IR4xOR2, which is able, to a certain degree, to alleviate catastrophic forgetting.
}

The best model, RepairLLaMA, fine-tuned with the code representation IR4xOR2, plausibly repairs 195 (+64) Defects4J bugs than the baseline, exactly repairs 124 (+72), syntactically correctly repairs 125 bugs (+55), and semantically correctly repairs 144 bugs (+61) bugs, respectively.
This highest performance is due to the presence of the buggy code in the input, which gives the model unique ingredients for understanding the problem and generating the patch.

To strengthen the external validity of our analysis beyond Defects4J, we perform the same experiment on HumanEval-Java \new{and GitBug-Java}.
On HumanEval-Java, RepairLLaMA also achieves better repair effectiveness than the baseline models (75 vs. 71 exact match, 82 vs. 81 AST match, and 109 vs. 103 semantic match).
\new{On GitBug-Java, RepairLLaMA also beats the baseline models (15 vs. 11 exact match, 16 vs. 12 AST match, and 20 vs. 13 semantic match).
These results are consistent with the observations made on Defects4J.}

\new{
Now, we discuss the statistical significance of our results.
All exact p-values are provided as supplemental material \cite{repairllama-statistical-test}.
We observe that there is a statistically significant difference between the golden model RepairLLaMA (IR4xOR2) and all other models, with very low p-values ($p \leq 1.27e-04$), with the exception of IR2xOR2 for which the test fails to reject the null hypothesis, this case is further discussed below.
These results completely validate the importance of the task-specific signals embedded in our best-performing code representations.
}

Third, we discuss the \new{impact of fine-tuning in aligning} the repair-specific code representation and the pre-training objectives.
IR2 and IR4 are input representations that include the same signals: the original function, fault localization, and the original buggy code.
The key difference is that IR4 uses the same \infilling scheme seen during pre-training, while providing the original buggy code in comments, while IR2 places two comments as delimiters of the buggy region.
\new{
We see a marginal performance gap between both representations (e.g., 144 vs. 139 correctly fixed Defects4J bugs) that is not statistically significant ($p \approx 0.545$) \cite{repairllama-statistical-test}.
We hypothesize that, while IR2 does not implicitly use the \infilling scheme, fine-tuning effectively maps the delimiting comments of IR2 to the \texttt{<FILL\_ME>} token used by IR4 and the pre-trained \infilling scheme.
}

\begin{figure}[t]
    \centering
    \includegraphics[width=\columnwidth]{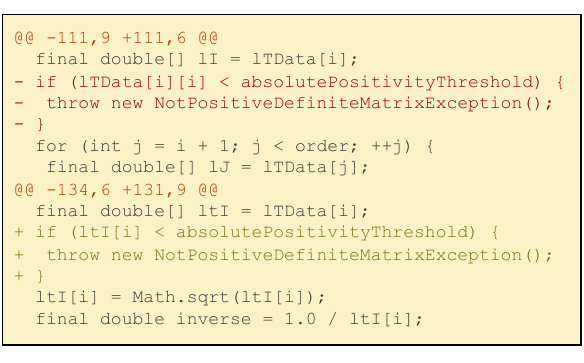}
    \caption{Exact match patch generated by RepairLLaMA for Math-86 from Defects4J v2. In this multi-location bug, RepairLLaMA is able to fix two distant buggy locations.}
    \label{fig:math-86}
\end{figure}

In addition to comparing performance, we explore the effectiveness of region-based representations in addressing multi-location bugs.
Our findings reveal that RepairLLaMA correctly repairs \new{51} instances of such bugs.
For example, RepairLLaMA correctly fixes a complex multi-chunk bug, Math-86, from the Defects4J v2, as illustrated in \autoref{fig:math-86}.
Math-86 presents two error sections that require simultaneous attention and correction: 1) the removal of an if block that throws an exception, and 2) the introduction of a new if condition.
Note that these two sections have more than 20 lines of distance between each other, showing that RepairLLaMA can fix bugs where the multiple edit locations are far away from each other.
To summarize, our experiments give clear evidence that RepairLLaMA's representation enables the repair of a wide range of multi-location bugs.

\begin{figure}[t]
    \centering
    \includegraphics[width=\columnwidth]{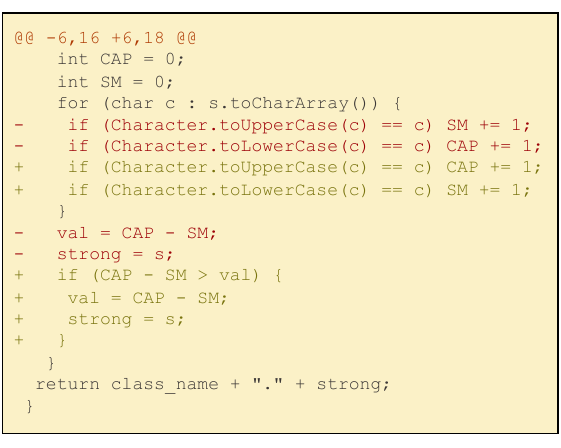}
    \caption{Exact match patch generated by RepairLLaMA for \textit{STRONGEST\_EXTENSION} from HumanEval-Java. In this multi-location bug, RepairLLaMA modifies two if blocks and encapsulates two other statements in a new if block.}
    \label{fig:strongest-extension}
\end{figure}

\autoref{fig:strongest-extension} shows another example of an exact match multi-location patch generated by RepairLLaMA, which is for a HumanEval-Java bug \textit{STRONGEST\_EXTENSION}.
RepairLLaMA first swaps the statements conditioned by the two existing if conditions, understanding that the wrong counters are being incremented in each case.
Then, RepairLLaMA conditions the two statements updating the current values of \textit{val} and \textit{strong} only if the difference between the counters is greater than the already existing solution.
Both \autoref{fig:strongest-extension} and \autoref{fig:math-86} show the effectiveness of RepairLLaMA in repairing multi-location bugs, thanks to tailored code representations that represent fault localization information in a realistic manner.

\begin{tcolorbox}
    \textbf{Answer to RQ1:}
    Our results demonstrate the importance of designing specific code representations for fine-tuning LLMs for APR.
    Naive representations such as full function are suboptimal, whether on the input or the output side of the model.
    Our experiments show that the best code representation pair is IR4xOR2, \new{which is due to leveraging two essential signals specific to the program repair task at hand (fault localization and the original buggy code)}.
    The RepairLLaMA model, fine-tuned with IR4xOR2, correctly repairs 144 Defects4J bugs, 109 HumanEval-Java bugs, \new{and 20 GitBug-Java bugs}, which is state of the art. This significant performance demonstrates the need for curated code representations in automated program repair. While the community focuses a lot on prompt engineering, our original experimental results encourage alternative research on domain-specific, expert code representations per downstream task in software engineering.
\end{tcolorbox}

\subsection{Results of RQ2 (Parameter-Efficient Fine-Tuning vs. Full Fine-Tuning)}

\begin{table*}
\centering
\caption{\andre{Repair effectiveness of parameter-efficient fine-tuned models compared with fully fine-tuned models (RQ2). RepairLLaMA, trained with parameter-efficient fine-tuning, outperforms its variants on Defects4J, HumanEval-Java, and GitBug-Java.}}
\label{tab:rq2_results}
\resizebox{\linewidth}{!}{%
\begin{tblr}{
  cells = {r},
  row{1} = {c},
  row{2} = {c},
  cell{1}{1} = {r=2}{},
  cell{1}{2} = {r=2}{},
  cell{1}{3} = {c=4}{},
  cell{1}{7} = {c=4}{},
  cell{1}{11} = {c=4}{},
  cell{3}{1} = {c},
  cell{3}{2} = {c},
  cell{4}{1} = {c},
  cell{4}{2} = {c},
  cell{5}{1} = {c},
  cell{5}{2} = {c},
  cell{6}{1} = {c},
  cell{6}{2} = {c},
  cell{7}{1} = {c},
  cell{7}{2} = {c},
  cell{8}{1} = {c},
  cell{8}{2} = {c},
  vlines,
  hline{1,3-9} = {-}{},
  hline{2} = {3-14}{},
}
{\textbf{Base}\\\textbf{Model}} & {\textbf{Fine-Tuning}\\\textbf{Method}} & {\textbf{Defects4J v2}\\\textbf{(488 bugs)}} &                &              &                   & {\textbf{HumanEval-Java}\\\textbf{(162 bugs)}} &                &              &                   & {\textbf{GitBug-Java}\\\textbf{(90 bugs)}} &                &              &                   \\
                                &                                         & Plausible                                    & {Exact\\Match} & {AST\\Match} & {Semantic\\Match} & Plausible                                      & {Exact\\Match} & {AST\\Match} & {Semantic\\Match} & Plausible                                  & {Exact\\Match} & {AST\\Match} & {Semantic\\Match} \\
codellama-7b                    & -                                       & 131                                          & 52             & 70           & 83                & 107                                            & 71             & 81           & 103               & 17                                         & 8              & 9            & 12                \\
codellama-7b                    & FFT                                     & 146                                          & 66             & 84           & 98                & 109                                            & 74             & \textbf{83}  & 100               & 21                                         & 11             & 11           & 13                \\
codellama-7b                    & LoRA                                    & \textbf{195}                                 & \textbf{124}   & \textbf{125} & \textbf{144}      & \textbf{118}                                   & \textbf{75}    & 82           & \textbf{109}      & \textbf{25}                                & \textbf{15}    & \textbf{16}  & \textbf{20}       \\
deepseek-coder-6.7b             & Base                                    & 147                                          & 84             & 88           &  104              & 113                                            & 68             & 83           & 110               & 21                                         & 11             & 11           & 13                \\
deepseek-coder-6.7b             & FFT                                     & \textbf{185}                                 & \textbf{116}   & \textbf{123} & \textbf{138}  & 129                                            & 96             & 101          & 119               & \textbf{26}                                & \textbf{15}    & \textbf{15}  & \textbf{19}   \\
deepseek-coder-6.7b             & LoRA                                    & 181                                          & 110            & 113          & 128               & \textbf{134}                                   & \textbf{99}    & \textbf{107} & \textbf{124}        & 19                                         & 12             & 12           & 13                
\end{tblr}
}
\end{table*}

In RQ2, we study how parameter-efficient fine-tuning compares against basic full-parameter fine-tuning.
Recall most of the closely related work in APR \cite{jiang2023impact, wang2023rap, shirafuji2023program, xia2023plastic, yuan2022circle, CURE-icse21} use full-parameter fine-tuning.
On the contrary, the key novelty of RepairLLaMA is that all models are fine-tuned using LoRA, a parameter-efficient fine-tuning technique that optimizes only a small adapter (approx. 4M parameters) instead of the whole LLM (approx. 7B parameters, a reduction of approx. 1600x).
This allows the model to 1) be fine-tuned with less GPU memory, and 2) potentially reduce overfitting.
In RQ2, we compare RepairLLaMA, built with the best code representation in RQ1, with its full-parameter fine-tuning version.
\new{Additionally, we also study the same fine-tuning methods with a different base model, \textit{deepseek-coder-6.7b}, to increase external validity.}

\autoref{tab:rq2_results} presents the results of RQ2.
The table reads as follows.
\andre{The first column presents the base model.
The second column presents the fine-tuning model.
The third to fifth meta-columns show the repair effectiveness results on, respectively, Defects4J v2, HumanEval-Java, and GitBug-Java.}

The results show that RepairLLaMA with parameter-efficient fine-tuning clearly outperforms both the baseline and its full-parameter fine-tuning version.
In Defects4J, RepairLLaMA plausibly repairs 49 (195 vs. 146) bugs more and correctly repairs, according to semantic match, 46 (144 vs. 98) more than the fully fine-tuned model.
When considering AST Match, RepairLLaMA also repairs 41 (125 vs. 84) more bugs.
In HumanEval-Java \new{and GitBug-Java}, although the improvement is smaller, RepairLLaMA still outperforms all full fine-tuning baselines.
\new{We find a statistical significance between the golden model IR4xOR2 and the two baselines ($p \leq 1.29e-03$, see \cite{repairllama-statistical-test}).}

The gain in performance is clear, and it is actually a double win because RepairLLaMA requires fewer resources.
Possibly, the fully fine-tuned models suffer from overfitting due to the limited fine-tuning data, while parameter-efficient fine-tuning helps prevent overfitting since it only requires optimizing a small part of the network weights.
The model size constraints of LoRA appear to act as implicit regularizers in this experiment.
\new{These results confirm the findings of \citeauthor{fu2023effectiveness} \cite{fu2023effectiveness}: parameter-efficient fine-tuning helps mitigate overfitting by leveraging trainable parameter sparsity and reducing the generalization error bound. Yet, a balance is needed to prevent underfitting due to insufficient trainable parameters.}

This phenomenon is not observed for \textit{deepseek-coder-6.7b}.
For this model, fine-tuning clearly improves over the base model.
The best fine-tuned model, for each benchmark, repairs 34 (138 vs. 104) more Defects4J, 24 (107 vs. 83) more HumanEval-Java, and 6 (19 vs. 13) more GitBug-Java bugs than the base model.
The full fine-tuned version outperforms LoRA in Defects4J and GitBug-Java, while the LoRA version outperforms in HumanEval-Java.
\newnew{
There is a statistically significant difference between both the fully fine-tuned ($p \approx 2.72e-05$) and LoRA versions ($p \approx 4.21e-04$) of deepseek-coder-6.7b when compared with the baseline model, meaning that both methods perform a statistically significant improvement over the baseline model.
}
These results suggest that the effect of parameter-efficient fine-tuning varies across models. It can under- or over-perform depending on the model and benchmark, always with the advantage of smaller resource requirements.

When compared with the best fully fine-tuned model from \citeauthor{jiang2023impact} \cite{jiang2023impact}, which keeps the LLMs' original pre-training format but includes the buggy lines as comments, \newnew{RepairLLaMA correctly repairs 25 (95 vs. 69) Defects4J bugs more, when applied to the same dataset considered in \citeauthor{jiang2023impact} and 39 (109 vs. 70) HumanEval-Java bugs more.
In addition, RepairLLaMA correctly repairs 49 more Defects4J multi-hunk bugs.}
Notably, RepairLLaMA achieves these results without access to perfect bug localization information, which is assumed and incorporated in \cite{jiang2023impact}.

To further evaluate RepairLLaMA, we also compare it with RAP-Gen \cite{wang2023rap}, the state-of-the-art program repair approach achieved by full-parameter fine-tuning CodeT5.
\new{RAP-Gen builds code representations by augmenting the original input of buggy code with retrieved bug-fix pairs.}
RAP-Gen correctly repairs 125 Defects4J v2 bugs, according to the authors' own manual verification, when generating 100 patches per bug \new{and providing perfect line-level fault localization.}
\new{In contrast, RepairLLaMA correctly repairs 19 more Defects4J v2 bugs (144 vs. 125), while generating only 10 patches per bug (10x less than RAP-Gen) and operating under more realistic fault localization.}
This clearly demonstrates that RepairLLaMA improves the state-of-the-art of fine-tuning for program repair.

\sen{Figure~\ref{fig:math-86} shows a multi-location bug from Math-86 (Defects4J v2) where RepairLLaMA successfully generates an exact match patch by fixing two distant buggy locations simultaneously. This example clearly demonstrates the limitations of both comparison approaches: Jiang et al.'s approach can only handle single-chunk bugs, and its representation is inadequate for addressing bugs that span multiple locations, as shown in this example. RAG-Gen, while theoretically capable, requires perfect fault localization information, which is impractical in real-world scenarios. In contrast, RepairLLaMA overcomes these limitations by tailoring effective code representation that can handle complex multi-location bugs without requiring perfect fault localization information.}

\begin{tcolorbox}
    \textbf{Answer to RQ2:}
    \andre{The efficacy of fine-tuning is clear, yet depends on the base model.
    RepairLLaMA's parameter-efficient fine-tuning outperforms full fine-tuning of \textit{codellama-7b}, with external validity over three benchmarks, while results for \textit{deepseek-coder-6.7b} show similar improvements regardless of the fine-tuning method.}
    Overall, our paper is the first to demonstrate the benefits of parameter-efficient fine-tuning in the context of automated program repair, achieving strong state-of-the-art results.
    Beyond program repair, we note that parameter-efficient fine-tuning is feasible within academic labs' budget and hardware, while still working with powerful multi-million dollar trained large models.
\end{tcolorbox}

\subsection{Results of RQ3 (RepairLLaMA vs. ChatGPT-based APR)}
\label{sec:results:rq3}

In RQ3, we compare RepairLLaMA with state-of-the-art zero-shot ChatGPT-based program repair.
ChatGPT-based program repair differs from RepairLLaMA since it does not involve fine-tuning LLMs on task-specific datasets.
Instead, it involves designing effective prompt strategies to instruct a powerful general-purpose LLM like GPT-4.

\begin{table*}
\centering
\caption{RepairLLaMA's effectiveness compared with state-of-the-art ChatGPT-based APR techniques (RQ3). RepairLLaMA is more effective in finding correct and plausible patches in Defects4J \new{and GitBug-Java}, incl. against the strong baseline of GPT-4.}
\label{tab:chatgpt}
\resizebox{\linewidth}{!}{%
\begin{tblr}{
  cells = {r},
  row{1} = {c},
  row{2} = {c},
  cell{1}{1} = {r=2}{},
  cell{1}{2} = {c=4}{},
  cell{1}{6} = {c=4}{},
  cell{1}{10} = {c=4}{},
  cell{3}{1} = {c},
  cell{4}{1} = {c},
  cell{5}{1} = {c},
  vlines,
  hline{1,4-6} = {-}{},
  hline{2-3} = {1-13}{},
}
\textbf{Model}                  & {\textbf{Defects4J v2}\\\textbf{(488 bugs)}} &                &              &                   & {\textbf{HumanEval-Java}\\\textbf{(162 bugs)}} &                &              &                   & {\textbf{GitBug-Java}\\\textbf{(90 bugs)}} &                &              &                   \\
                                & Plausible                                    & {Exact\\Match} & {AST\\Match} & {Semantic\\Match} & Plausible                                      & {Exact\\Match} & {AST\\Match} & {Semantic\\Match} & Plausible                                  & {Exact\\Match} & {AST\\Match} & {Semantic\\Match} \\
GPT-3.5                         & 71                                           & 23             & 33           & 45                & 107                                            & 50             & 63           & 97                & 9                                          & 7              & 7            & 8                 \\
GPT-4                           & 119                                          & 47             & 60           & 72                & \textbf{124}                                   & 64             & 74           & \textbf{116}      & 14                                         & 6              & 7            & 10                \\
\textbf{RepairLLaMA (IR4xOR2)} & \textbf{195}                                 & \textbf{124}   & \textbf{125} & \textbf{144}      & 118                                            & \textbf{75}    & \textbf{82}  & 109               & \textbf{25}                                         & \textbf{15}             & \textbf{16}           & \textbf{20}                
\end{tblr}
}
\end{table*}

\autoref{tab:chatgpt} shows the repair effectiveness of RepairLLaMA compared with ChatGPT-based APR techniques.
The first column indicates the model name.
The second, and third show the results on two different benchmarks.
Each benchmark is evaluated following the four patch assessment metrics described in \autoref{sec:exp:assess}.

Our results show that RepairLLaMA is the most effective model.
GPT-3.5 and GPT-4 plausibly fix 64\% and 39\% less Defects4J v2 bugs than RepairLLaMA, respectively.
RepairLLaMA correctly fixes twice as many bugs as GPT4, when considering any of the correct fix metrics (i.e. Exact/AST/Semantic Match).
GPT-4 correctly repairs more HumanEval-Java bugs than RepairLLaMA according to semantic match, which may be due to overfitting to HumanEval. RepairLLaMA correctly repairs more according to both Exact/AST Match metrics.

\new{
RepairLLaMA also plausibly and correctly fixes more GitBug-Java bugs than both GPT-3.5 and GPT-4, showing external validity.
GitBug-Java exclusively contains bugs from 2023, meaning that all bugs occurred after the training data cutoff date of all three models.
Moreover, we observe a statistically significant difference between RepairLLaMA and GPT-3.5/GPT-4 ($p \leq 4.10e-08$, see \cite{repairllama-statistical-test}).}
Recall that both GPT-3.5 and GPT-4 are larger than the RepairLLaMA model by several orders of magnitude, making RepairLLaMA's specialized performance remarkable.

To conclude, our experimental results demonstrate the power of specializing an LLM for APR over zero-shot learning of a foundational model.
A smaller model, trained with a parameter-efficient fine-tuning technique, is more effective than a large general-purpose LLM.
Overall, RepairLLaMA beats the strong baseline of GPT-4 on the hardest benchmark Defects4J.

\begin{tcolorbox}
    \textbf{Answer to RQ3:}
    RepairLLaMA beats GPT-4, thanks to the combination of appropriate code representations and parameter-efficient fine-tuning.
    Our experiments demonstrate the need for specialized representations and specialized training for program repair. The RepairLLaMA program repair adapter is more powerful than ChatGPT for fixing bugs in Java.
\end{tcolorbox}

\section{Discussion}

\subsection{Sampling Candidate Patches}

One key aspect of any program repair approach is the number of generated candidate patches.
Some recent works generate hundreds and even thousands of patches for a single bug \cite{xia2023keep, xia2022less, wei2023copiloting, wang2023rap, xiang2024far}.
However, the cost of evaluating such a large number of candidate patches has been largely overlooked.
Recall that to evaluate the plausibility of each candidate patch, one must run the test cases, which is expensive and even overcomes the one-time cost of fine-tuning.
In contrast to this trend, RepairLLaMA achieves state-of-the-art results while generating only 10 candidate patches per bug.
This shows that 
1) RepairLLaMA natively priorities the best patches in the top-10 list, 
2) RepairLLaMA minimizes the resources that are required in an end-to-end repair pipeline that includes plausibility checking.

\subsection{\new{Fault Localization with Code Regions}}

\new{
In most recent program repair articles, effectiveness measurement assumes perfect fault localization (FL).
Such an assumption, while convenient for research purposes, is unrealistic in real-world scenarios since current FL techniques cannot reliably locate all faulty code lines with high accuracy.
This is particularly true in cases where the buggy lines are spread across different locations since there are virtually no multi-location fault localization algorithms.
To enhance practical applicability, we propose in this paper an alternative, relaxed assumption: instead of making a perfect localization assumption, we only assume that one is able to identify a code region that encompasses all the buggy lines.
}

\new{
For a function with $N$ code lines, there are  maximum $\frac{N(N+1)}{2}$ code regions.
In other words, the faulty code region (i.e., the shortest region containing all buggy lines) can be identified with $O(N^2)$ complexity. This makes it computationally feasible to generate candidate regions by examining the different pairs of start and end lines, if $N$ is not too lage.
We implement \footnote{\url{https://github.com/ASSERT-KTH/repairllama/blob/main/src/fl/fl.py}} and analyze \footnote{\url{https://github.com/ASSERT-KTH/repairllama/blob/main/src/fl/analyse_fl.ipynb}} the proposed approach for all three considered benchmarks (Defects4Jv2, HumanEval-Java, GitBug-Java). 
We observe that the 75th percentile of the number of code regions in the buggy functions of each benchmark is approx. 8485 for GitBug-Java, 861 for Defects4J, and 91 for HumanEval-Java. This shows the feasibility of enumerating all possible regions for most single-function bugs of the considered benchmarks.
}

\new{
We believe that this novel strategy is realistically better: even if the localization is imperfect, as long as it can narrow down the search space to a region containing all buggy lines, a model such as RepairLLaMA would be effective in generating multi-location patches.
}

\new{
To sum up, our code region strategy 1) dramatically relaxes the assumption of perfect multi-line fault localization of related research, and 2) enables mor practical program repair.
}

\subsection{\new{Fine-Tuning vs. Prompt Engineering}}

\new{Fine-tuning and prompt engineering are two distinct strategies for utilizing LLMs in program repair.
Prompt engineering leverages the inherent capabilities of LLMs to generate bug fixes by steering the models with carefully crafted prompts, often including examples and detailed instructions.
Prompt engineering requires no additional training and can be rapidly used for new tasks.
However, its reliance on ad hoc prompt engineering introduces variability (e.g., what works for one model might not work for another), bounds the amount of task-specific information that can be provided to the model to its context window, and is usually dependent on large, proprietary models.
}

\new{
Fine-tuning, as we do in this paper, adapts the LLM by learning domain-specific tokens, structures and signals.
In our program repair context, this means learning the relationship between buggy and fixed code by directly changing the model's weights.
This results in consistent, high-quality fixes, even with smaller models.
Although fine-tuning requires task-specific datasets and more computation resources, methods like LoRA help mitigate these costs.
While both approaches have advantages, our results confirm the effectiveness of fine-tuning in achieving good performance in program repair.
}

\subsection{Threats to Validity}
The primary internal threat lies in the potential data leakage during the pre-training phase of LLMs. 
LLMs are pre-trained on vast corpora scrapped from the web, potentially including the same data used for testing, endangering the reliability of experimental results.
To mitigate this threat, we assess all models on two recent benchmarks specifically designed to address the data leakage issue \cite{ramos2024large}, HumanEval-Java \cite{jiang2023impact} \new{and GitBug-Java \cite{silva2024gitbug}}.

Another internal threat pertains to data leakage during the fine-tuning process of LLMs, since both our fine-tuning dataset, Megadiff, and Defects4J contain samples from GitHub from the same period of time.
To address this threat, we meticulously compare the samples in our fine-tuning dataset, Megadiff, with those in Defects4J.
We found no identical samples shared by both datasets.
However, it is worth noting that there are three samples (Math-28, Math-44, and JacksonDatabind-82) whose patch includes a function also found in Megadiff samples.
To mitigate this threat, we exclude these three Defects4J samples from the evaluation of our fine-tuned models.
\new{Moreover, both HumanEval-Java and GitBug-Java have been constructed after Megadiff and, as such, are non-overlapping.}

\new{The main external threats are two-fold: 1)} the focus on a single programming language, as our results may not generalize to other languages, \new{and 2) the focus on a single LLM.}

To mitigate the first threat, we evaluate on three benchmarks, including well-established Defects4J \cite{defects4j}.
Our core novelties are programming language agnostic and should generalize to arbitrary languages.

\new{Regarding the second threat, we argue that our approach is model agnostic.
The only model-specific feature we rely on is the infilling (aka fill-in-the-middle) feature.
This feature is commonly used in LLMs' training for code \cite{codestral, li2023starcoder, team2024codegemma, fried2022incoder}, and, as such, our approach is easily extensible to these models.
For the LLMs that are not trained with this objective, we note that it is also possible to fine-tune them with the proposed representations.}

\section{Related Work}

\subsection{Large Language Models for Program Repair}

\emph{Fine-Tuning. }
Several works \cite{CURE-icse21, yuan2022circle, zhang2022coditt5, zirak2022improving, jin2023inferfix, xia2023plastic, wang2023rap, shirafuji2023program, paulenhancing, jiang2023impact, hao2023enhancing, Hossain2024} have proposed fine-tuning large language models for the program repair task.
Notably, \citeauthor{jiang2023impact} \cite{jiang2023impact} specifically study the impact of fine-tuning LLMs for program repair, reporting improvements lower than ours while using naive full-parameter fine-tuning.
\citeauthor{huang2023empirical} \cite{huang2023empirical} also study different aspects of fine-tuning LLMs for program repair, including code representations and evaluation metrics.
While they report state-of-the-art performance, it is achieved under the unrealistic assumption of perfect multi-line fault localization, which RepairLLaMA does not assume.
\citeauthor{yang2024multi} \cite{yang2024multi} propose MORepair, an approach to fine-tuning CodeLlama for program repair with multiple objectives, including LLM-generated guidance text in the novel fine-tuning objective.
MORepair takes inspiration from RepairLLaMA in using parameter-efficient fine-tuning to train a repair adapter, comparing MORepair with RepairLLaMA on two new benchmarks built from HumanEval \cite{chen2021evaluating}.

Overall, our work distinguishes itself from related work in three key aspects.
First, we have designed and evaluated several code representations in RepairLLaMA, tailored to fine-tuning LLMs for program repair, which incorporate fault localization signals under realistic assumptions.
This is different from previous work (e.g., \cite{wang2023rap, huang2023empirical}) which assumes perfect multi-line fault localization.
Second, RepairLLaMA is the first to employ LoRA to fine-tune LLMs for program repair, demonstrating that parameter-efficient fine-tuning can surpass full-parameter fine-tuning while reducing computational requirements.
Third, unlike some previous work that generates hundreds, even thousands of patches for each bug, our best model, RepairLLaMA, improves state-of-the-art performance on Defects4J and HumanEval-Java with a budget of just 10 patches per bug, demonstrating the laser-style focus of the trained program repair adapter.

\emph{Prompting and Agents. }
Many works \cite{xia2022less, xia2022practical, ahmed2023majority, huang2023chain, deligiannis2023fixing, wei2023copiloting, zhang2023critical, xiang2024far} directly use LLMs for program repair without any fine-tuning.
The core of these works is prompting: they design and evaluate different prompting strategies to provide repair information to the model.
\new{Notably, SRepair \cite{xiang2024far} uses a dual-LLM approach that first generates natural language repair suggestions by providing the buggy function, failing tests, and error messages in a chain-of-thought prompt to an LLM.
Then, SRepair prompts a smaller LLM to generate up to 500 patches per bug based on the high-level suggestions.}
Recently, related work \cite{xia2023keep, yang2024swe, zhang2024autocoderover, bouzenia2024repairagent} has also proposed agents for program repair.
The core novelty is to prompt LLMs iteratively, with intermediate steps for tool execution and incorporation of their feedback in the prompting loop.

Our work is fundamentally different from this related work: 1) we do not do prompt engineering, and 2) we do not iterative repair.
RepairLLaMA crystallizes all the repair knowledge in the model weights obtained by parameter-efficient fine-tuning. The concept of program repair adapter is entirely different and new compared to prompting.
Remarkably, RepairLLaMa achieves strong performance while being zero-shot without access to external tools and feedback.
We consider our findings to be the foundation of future work in fine-tuning for software agents.

\subsection{Code Representations for Program Repair}
Several code representations for program repair have been proposed by related work \cite{SEQUENCER, deepfix, namavar2022controlled, abhinav2021repairnet, chen2022neural, yasunaga2020graph, chen2021plur, ye2022selfapr, muennighoff2023octopack}.
Notably, \citeauthor{namavar2022controlled} \cite{namavar2022controlled} investigate the impact of different code representations for program repair for a restricted set of bug classes.
Differently, our work targets a larger spectrum of bugs, including multi-location bugs.

Overall, our work distinguishes itself from preceding research in code representation in three dimensions.
First, we design code representations that are aligned with the pre-training data and objectives, enabling the RepairLLaMA to well utilize the pre-learned knowledge.
Second, our code representations are designed to support a large spectrum of bugs, including multi-location bugs, which is one frontier of program repair.
Third, RepairLLaMA's pipeline and evaluation are not constructed under the unrealistic perfect fault localization for multi-location bugs.

\subsection{Parameter-Efficient Fine-Tuning in SE}

Parameter-efficient fine-tuning is a relatively under-explored area in Software Engineering.
\citeauthor{wang2023one} \cite{wang2023one} explore parameter-efficient fine-tuning techniques for specializing LLMs for code search and code summarization, finding that parameter-efficient fine-tuning outperforms in-context learning.
\citeauthor{weyssow2023exploring} \cite{weyssow2023exploring} confirm the dominance of parameter-efficient fine-tuning techniques over zero-shot learning in code generation, while \citeauthor{wang2022no} \cite{wang2022no} find that prompt-tuning outperforms traditional fine-tuning methods in code summarization.
CodePrompt \cite{choi2023codeprompt} proposes corpus-specific prompt templates similar to adaptations and boosts code generation performance.
\citeauthor{shi2023towards} \cite{shi2023towards} propose a parameter-efficient fine-tuning technique for code related tasks that selectively freezes layers of the model.
\citeauthor{zou2023comprehensive} \cite{zou2023comprehensive} study the effectiveness of five parameter-efficient fine-tuning methods in four software engineering tasks, finding them competitive with full-parameter fine-tuning.
Lastly, \citeauthor{liu2024delving} \cite{liu2024delving} study the performance of Adapter Tuning and LoRA in different software engineering tasks, including cross-lingual and low-resource scenarios, finding PEFT to outperform or achieve comparable results when compared with full-parameter fine-tuning.

Overall, we are the first to employ and evaluate LoRA to fine-tune LLMs for program repair. RepairLLaMA's effectiveness calls for more work in parameter-efficient fine-tuning in program repair and related tasks such as overfitting detection.

\section{Conclusion}

In this paper, we have proposed RepairLLaMA, a novel program repair approach that combines parameter-efficient fine-tuning with program repair-specific code representations.
RepairLLaMA's code representations are unique in incorporating repair signals, such as fault localization, under realistic assumptions, and in aligning with pre-training data and objectives.
To validate RepairLLaMA, we perform a series of extensive experiments on three benchmarks: Defects4J, HumanEval-Java, and GitBug-Java.
Our results clearly validate our core design decisions, with RepairLLaMA correctly fixing 144 Defects4J, 109 HumanEval-Java bugs, \new{and 20 GitBug-Java bugs}, outperforming strong baselines, incl. GPT-3.5 and GPT-4.
RepairLLaMA opens an avenue for research on different kinds of efficient fine-tuning for program repair.

\section{Acknowledgments}

The computations/data handling were enabled by the supercomputing resources Berzelius-2023-175 and Berzelius-2024-336 provided by National Supercomputer Centre at Linköping University and the Knut and Alice Wallenberg foundation.
This work was partially supported by the Wallenberg AI, Autonomous Systems and Software Program (WASP) funded by the Knut and Alice Wallenberg Foundation.

\balance
\bibliography{IEEEabrv,ref}

% Generated by IEEEtranN.bst, version: 1.14 (2015/08/26)
\begin{thebibliography}{70}
\providecommand{\natexlab}[1]{#1}
\providecommand{\url}[1]{#1}
\csname url@samestyle\endcsname
\providecommand{\newblock}{\relax}
\providecommand{\bibinfo}[2]{#2}
\providecommand{\BIBentrySTDinterwordspacing}{\spaceskip=0pt\relax}
\providecommand{\BIBentryALTinterwordstretchfactor}{4}
\providecommand{\BIBentryALTinterwordspacing}{\spaceskip=\fontdimen2\font plus
\BIBentryALTinterwordstretchfactor\fontdimen3\font minus \fontdimen4\font\relax}
\providecommand{\BIBforeignlanguage}[2]{{%
\expandafter\ifx\csname l@#1\endcsname\relax
\typeout{** WARNING: IEEEtranN.bst: No hyphenation pattern has been}%
\typeout{** loaded for the language `#1'. Using the pattern for}%
\typeout{** the default language instead.}%
\else
\language=\csname l@#1\endcsname
\fi
#2}}
\providecommand{\BIBdecl}{\relax}
\BIBdecl

\bibitem[Monperrus(2018)]{monperrus2018automatic}
M.~Monperrus, ``Automatic software repair: A bibliography,'' \emph{ACM Computing Surveys (CSUR)}, vol.~51, no.~1, pp. 1--24, 2018.

\bibitem[Le~Goues et~al.(2021)Le~Goues, Pradel, Roychoudhury, and Chandra]{le2021automatic}
C.~Le~Goues, M.~Pradel, A.~Roychoudhury, and S.~Chandra, ``Automatic program repair,'' \emph{IEEE Software}, vol.~38, no.~4, pp. 22--27, 2021.

\bibitem[Chen et~al.(2019)Chen, Kommrusch, Tufano, Pouchet, Poshyvanyk, and Monperrus]{chen2019sequencer}
Z.~Chen, S.~Kommrusch, M.~Tufano, L.-N. Pouchet, D.~Poshyvanyk, and M.~Monperrus, ``Sequencer: Sequence-to-sequence learning for end-to-end program repair,'' \emph{IEEE Transactions on Software Engineering}, vol.~47, no.~9, pp. 1943--1959, 2019.

\bibitem[Zhu et~al.(2021)Zhu, Sun, Xiao, Zhang, Yuan, Xiong, and Zhang]{Recoder}
Q.~Zhu, Z.~Sun, Y.-a. Xiao, W.~Zhang, K.~Yuan, Y.~Xiong, and L.~Zhang, ``A syntax-guided edit decoder for neural program repair,'' in \emph{Proceedings of the 29th ACM Joint Meeting on European Software Engineering Conference and Symposium on the Foundations of Software Engineering}, ser. ESEC/FSE 2021.\hskip 1em plus 0.5em minus 0.4em\relax ACM, 2021, p. 341–353.

\bibitem[Jiang et~al.(2021{\natexlab{a}})Jiang, Lutellier, and Tan]{CURE-icse21}
N.~Jiang, T.~Lutellier, and L.~Tan, ``Cure: Code-aware neural machine translation for automatic program repair,'' in \emph{Proceedings of the ACM/IEEE 43rd International Conference on Software Engineering}, 2021.

\bibitem[Ye et~al.(2022)Ye, Martinez, Luo, Zhang, and Monperrus]{ye2022selfapr}
H.~Ye, M.~Martinez, X.~Luo, T.~Zhang, and M.~Monperrus, ``Selfapr: Self-supervised program repair with test execution diagnostics,'' in \emph{Proceedings of the 37th IEEE/ACM International Conference on Automated Software Engineering}, 2022, pp. 1--13.

\bibitem[Xia and Zhang(2022)]{xia2022less}
C.~S. Xia and L.~Zhang, ``Less training, more repairing please: revisiting automated program repair via zero-shot learning,'' in \emph{Proceedings of the 30th ACM Joint European Software Engineering Conference and Symposium on the Foundations of Software Engineering}, 2022, pp. 959--971.

\bibitem[Jiang et~al.(2023)Jiang, Liu, Lutellier, and Tan]{jiang2023impact}
N.~Jiang, K.~Liu, T.~Lutellier, and L.~Tan, ``Impact of code language models on automated program repair,'' in \emph{2023 IEEE/ACM 45th International Conference on Software Engineering (ICSE)}, 2023, pp. 1430--1442.

\bibitem[Xia et~al.(2023)Xia, Ding, and Zhang]{xia2023plastic}
C.~S. Xia, Y.~Ding, and L.~Zhang, ``The plastic surgery hypothesis in the era of large language models,'' in \emph{2023 38th IEEE/ACM International Conference on Automated Software Engineering (ASE)}.\hskip 1em plus 0.5em minus 0.4em\relax IEEE, 2023, pp. 522--534.

\bibitem[Wang et~al.(2023{\natexlab{a}})Wang, Wang, Joty, and Hoi]{wang2023rap}
W.~Wang, Y.~Wang, S.~Joty, and S.~C. Hoi, ``Rap-gen: Retrieval-augmented patch generation with codet5 for automatic program repair,'' in \emph{Proceedings of the 31st ACM Joint European Software Engineering Conference and Symposium on the Foundations of Software Engineering}, 2023, pp. 146--158.

\bibitem[Xia et~al.(2022)Xia, Wei, and Zhang]{xia2022practical}
C.~S. Xia, Y.~Wei, and L.~Zhang, ``Practical program repair in the era of large pre-trained language models,'' \emph{arXiv preprint arXiv:2210.14179}, 2022.

\bibitem[Yuan et~al.(2022)Yuan, Zhang, He, Fang, Hung, Hao, and Yin]{yuan2022circle}
W.~Yuan, Q.~Zhang, T.~He, C.~Fang, N.~Q.~V. Hung, X.~Hao, and H.~Yin, ``Circle: Continual repair across programming languages,'' in \emph{Proceedings of the 31st ACM SIGSOFT International Symposium on Software Testing and Analysis}, 2022, pp. 678--690.

\bibitem[Joshi et~al.(2023)Joshi, Sanchez, Gulwani, Le, Verbruggen, and Radi{\v{c}}ek]{joshi2023repair}
H.~Joshi, J.~C. Sanchez, S.~Gulwani, V.~Le, G.~Verbruggen, and I.~Radi{\v{c}}ek, ``Repair is nearly generation: Multilingual program repair with llms,'' in \emph{Proceedings of the AAAI Conference on Artificial Intelligence}, vol.~37, no.~4, 2023, pp. 5131--5140.

\bibitem[Xia and Zhang(2023)]{xia2023keep}
C.~S. Xia and L.~Zhang, ``Keep the conversation going: Fixing 162 out of 337 bugs for \$0.42 each using chatgpt,'' \emph{arXiv preprint arXiv:2304.00385}, 2023.

\bibitem[Yang et~al.(2024{\natexlab{a}})Yang, Jimenez, Wettig, Lieret, Yao, Narasimhan, and Press]{yang2024swe}
J.~Yang, C.~E. Jimenez, A.~Wettig, K.~Lieret, S.~Yao, K.~Narasimhan, and O.~Press, ``Swe-agent: Agent-computer interfaces enable automated software engineering,'' \emph{arXiv preprint arXiv:2405.15793}, 2024.

\bibitem[Zhang et~al.(2024)Zhang, Ruan, Fan, and Roychoudhury]{zhang2024autocoderover}
Y.~Zhang, H.~Ruan, Z.~Fan, and A.~Roychoudhury, ``Autocoderover: Autonomous program improvement,'' \emph{arXiv preprint arXiv:2404.05427}, 2024.

\bibitem[Bouzenia et~al.(2024)Bouzenia, Devanbu, and Pradel]{bouzenia2024repairagent}
I.~Bouzenia, P.~Devanbu, and M.~Pradel, ``Repairagent: An autonomous, llm-based agent for program repair,'' \emph{arXiv preprint arXiv:2403.17134}, 2024.

\bibitem[Chen et~al.(2021{\natexlab{a}})Chen, Hellendoorn, Lamblin, Maniatis, Manzagol, Tarlow, and Moitra]{chen2021plur}
Z.~Chen, V.~J. Hellendoorn, P.~Lamblin, P.~Maniatis, P.-A. Manzagol, D.~Tarlow, and S.~Moitra, ``Plur: A unifying, graph-based view of program learning, understanding, and repair,'' \emph{Advances in Neural Information Processing Systems}, vol.~34, pp. 23\,089--23\,101, 2021.

\bibitem[Lu et~al.(2023)Lu, Yu, Li, Yang, and Zuo]{lu2023llama}
J.~Lu, L.~Yu, X.~Li, L.~Yang, and C.~Zuo, ``Llama-reviewer: Advancing code review automation with large language models through parameter-efficient fine-tuning (practical experience report),'' \emph{arXiv preprint arXiv:2308.11148}, 2023.

\bibitem[Hu et~al.(2021)Hu, Wallis, Allen-Zhu, Li, Wang, Wang, Chen, et~al.]{hu2021lora}
E.~J. Hu, P.~Wallis, Z.~Allen-Zhu, Y.~Li, S.~Wang, L.~Wang, W.~Chen \emph{et~al.}, ``Lora: Low-rank adaptation of large language models,'' in \emph{International Conference on Learning Representations}, 2021.

\bibitem[Fu et~al.(2023)Fu, Yang, So, Lam, Bing, and Collier]{fu2023effectiveness}
Z.~Fu, H.~Yang, A.~M.-C. So, W.~Lam, L.~Bing, and N.~Collier, ``On the effectiveness of parameter-efficient fine-tuning,'' in \emph{Proceedings of the AAAI conference on artificial intelligence}, vol.~37, no.~11, 2023, pp. 12\,799--12\,807.

\bibitem[Just et~al.(2014)Just, Jalali, and Ernst]{defects4j}
R.~Just, D.~Jalali, and M.~D. Ernst, ``Defects4j: A database of existing faults to enable controlled testing studies for java programs,'' in \emph{Proceedings of the 2014 International Symposium on Software Testing and Analysis}.\hskip 1em plus 0.5em minus 0.4em\relax ACM, 2014, pp. 437--440.

\bibitem[Silva et~al.(2024)Silva, Saavedra, and Monperrus]{silva2024gitbug}
A.~Silva, N.~Saavedra, and M.~Monperrus, ``Gitbug-java: A reproducible benchmark of recent java bugs,'' in \emph{2024 IEEE/ACM 21st International Conference on Mining Software Repositories (MSR)}.\hskip 1em plus 0.5em minus 0.4em\relax IEEE, 2024, pp. 118--122.

\bibitem[Roziere et~al.(2023)Roziere, Gehring, Gloeckle, Sootla, Gat, Tan, Adi, Liu, Remez, Rapin, et~al.]{roziere2023code}
B.~Roziere, J.~Gehring, F.~Gloeckle, S.~Sootla, I.~Gat, X.~E. Tan, Y.~Adi, J.~Liu, T.~Remez, J.~Rapin \emph{et~al.}, ``Code llama: Open foundation models for code,'' \emph{arXiv preprint arXiv:2308.12950}, 2023.

\bibitem[Huang et~al.(2023{\natexlab{a}})Huang, Meng, Zhang, Liu, Wang, Li, and Zhang]{huang2023empirical}
K.~Huang, X.~Meng, J.~Zhang, Y.~Liu, W.~Wang, S.~Li, and Y.~Zhang, ``An empirical study on fine-tuning large language models of code for automated program repair,'' in \emph{2023 38th IEEE/ACM International Conference on Automated Software Engineering (ASE)}.\hskip 1em plus 0.5em minus 0.4em\relax IEEE Computer Society, 2023, pp. 1162--1174.

\bibitem[Ye and Monperrus(2023)]{ye2023iter}
H.~Ye and M.~Monperrus, ``Iter: Iterative neural repair for multi-location patches,'' \emph{arXiv preprint arXiv:2304.12015}, 2023.

\bibitem[Touvron et~al.(2023)Touvron, Lavril, Izacard, Martinet, Lachaux, Lacroix, Rozi{\`e}re, Goyal, Hambro, Azhar, et~al.]{touvron2023llama}
H.~Touvron, T.~Lavril, G.~Izacard, X.~Martinet, M.-A. Lachaux, T.~Lacroix, B.~Rozi{\`e}re, N.~Goyal, E.~Hambro, F.~Azhar \emph{et~al.}, ``Llama: Open and efficient foundation language models,'' \emph{arXiv preprint arXiv:2302.13971}, 2023.

\bibitem[Li et~al.(2023)Li, Allal, Zi, Muennighoff, Kocetkov, Mou, Marone, Akiki, Li, Chim, et~al.]{li2023starcoder}
R.~Li, L.~B. Allal, Y.~Zi, N.~Muennighoff, D.~Kocetkov, C.~Mou, M.~Marone, C.~Akiki, J.~Li, J.~Chim \emph{et~al.}, ``Starcoder: may the source be with you!'' \emph{arXiv preprint arXiv:2305.06161}, 2023.

\bibitem[Bavarian et~al.(2022)Bavarian, Jun, Tezak, Schulman, McLeavey, Tworek, and Chen]{bavarian2022efficient}
M.~Bavarian, H.~Jun, N.~Tezak, J.~Schulman, C.~McLeavey, J.~Tworek, and M.~Chen, ``Efficient training of language models to fill in the middle,'' \emph{arXiv preprint arXiv:2207.14255}, 2022.

\bibitem[Namavar et~al.(2022)Namavar, Nashid, and Mesbah]{namavar2022controlled}
M.~Namavar, N.~Nashid, and A.~Mesbah, ``A controlled experiment of different code representations for learning-based program repair,'' \emph{Empirical Software Engineering}, vol.~27, no.~7, p. 190, 2022.

\bibitem[Saha et~al.(2019)Saha, Saha, and Prasad]{hercules}
S.~Saha, R.~K. Saha, and M.~R. Prasad, ``Harnessing evolution for multi-hunk program repair,'' in \emph{Proceedings of the 41st International Conference on Software Engineering}, ser. ICSE '19, 2019, p. 13–24.

\bibitem[Lv et~al.(2023)Lv, Yang, Liu, Gao, Guo, and Qiu]{lv2023full}
K.~Lv, Y.~Yang, T.~Liu, Q.~Gao, Q.~Guo, and X.~Qiu, ``Full parameter fine-tuning for large language models with limited resources,'' \emph{arXiv preprint arXiv:2306.09782}, 2023.

\bibitem[Monperrus et~al.(2021)Monperrus, Martinez, Ye, Madeiral, Durieux, and Yu]{monperrus2021megadiff}
M.~Monperrus, M.~Martinez, H.~Ye, F.~Madeiral, T.~Durieux, and Z.~Yu, ``Megadiff: A dataset of 600k java source code changes categorized by diff size,'' 2021.

\bibitem[Jiang et~al.(2021{\natexlab{b}})Jiang, Lutellier, and Tan]{jiang2021cure}
N.~Jiang, T.~Lutellier, and L.~Tan, ``Cure: Code-aware neural machine translation for automatic program repair,'' in \emph{2021 IEEE/ACM 43rd International Conference on Software Engineering (ICSE)}.\hskip 1em plus 0.5em minus 0.4em\relax IEEE, 2021, pp. 1161--1173.

\bibitem[Chen et~al.(2021{\natexlab{b}})Chen, Tworek, Jun, Yuan, Pinto, Kaplan, Edwards, Burda, Joseph, Brockman, et~al.]{chen2021evaluating}
M.~Chen, J.~Tworek, H.~Jun, Q.~Yuan, H.~P. d.~O. Pinto, J.~Kaplan, H.~Edwards, Y.~Burda, N.~Joseph, G.~Brockman \emph{et~al.}, ``Evaluating large language models trained on code,'' \emph{arXiv preprint arXiv:2107.03374}, 2021.

\bibitem[Pawlak et~al.(2016)Pawlak, Monperrus, Petitprez, Noguera, and Seinturier]{spoon}
\BIBentryALTinterwordspacing
R.~Pawlak, M.~Monperrus, N.~Petitprez, C.~Noguera, and L.~Seinturier, ``Spoon: A library for implementing analyses and transformations of java source code,'' \emph{Softw. Pract. Exper.}, vol.~46, no.~9, p. 1155–1179, Sep. 2016. [Online]. Available: \url{https://doi.org/10.1002/spe.2346}
\BIBentrySTDinterwordspacing

\bibitem[Falleri et~al.(2014)Falleri, Morandat, Blanc, Martinez, and Monperrus]{falleri2014fine}
J.-R. Falleri, F.~Morandat, X.~Blanc, M.~Martinez, and M.~Monperrus, ``Fine-grained and accurate source code differencing,'' in \emph{Proceedings of the 29th ACM/IEEE international conference on Automated software engineering}, 2014, pp. 313--324.

\bibitem[Guo et~al.(2024)Guo, Zhu, Yang, Xie, Dong, Zhang, Chen, Bi, Wu, Li, et~al.]{guo2024deepseek}
D.~Guo, Q.~Zhu, D.~Yang, Z.~Xie, K.~Dong, W.~Zhang, G.~Chen, X.~Bi, Y.~Wu, Y.~Li \emph{et~al.}, ``Deepseek-coder: When the large language model meets programming--the rise of code intelligence,'' \emph{arXiv preprint arXiv:2401.14196}, 2024.

\bibitem[Fried et~al.(2022)Fried, Aghajanyan, Lin, Wang, Wallace, Shi, Zhong, Yih, Zettlemoyer, and Lewis]{fried2022incoder}
D.~Fried, A.~Aghajanyan, J.~Lin, S.~Wang, E.~Wallace, F.~Shi, R.~Zhong, W.-t. Yih, L.~Zettlemoyer, and M.~Lewis, ``Incoder: A generative model for code infilling and synthesis,'' \emph{arXiv preprint arXiv:2204.05999}, 2022.

\bibitem[Zhang et~al.(2023)Zhang, Zhang, Zhai, Fang, Yu, Sun, and Chen]{zhang2023critical}
Q.~Zhang, T.~Zhang, J.~Zhai, C.~Fang, B.~Yu, W.~Sun, and Z.~Chen, ``A critical review of large language model on software engineering: An example from chatgpt and automated program repair,'' \emph{arXiv preprint arXiv:2310.08879}, 2023.

\bibitem[rep()]{repairllama-statistical-test}
``Repairllama - statistical test,'' \url{https://github.com/ASSERT-KTH/repairllama/blob/main/src/patch_analysis/statistical_testing.md}, [Accessed 05-11-2024].

\bibitem[Shirafuji et~al.(2023)Shirafuji, Rahman, Amin, and Watanobe]{shirafuji2023program}
A.~Shirafuji, M.~M. Rahman, M.~F.~I. Amin, and Y.~Watanobe, ``Program repair with minimal edits using codet5,'' \emph{arXiv preprint arXiv:2309.14760}, 2023.

\bibitem[Wei et~al.(2023)Wei, Xia, and Zhang]{wei2023copiloting}
Y.~Wei, C.~S. Xia, and L.~Zhang, ``Copiloting the copilots: Fusing large language models with completion engines for automated program repair,'' \emph{arXiv preprint arXiv:2309.00608}, 2023.

\bibitem[Xiang et~al.(2024)Xiang, Xu, Kong, Wu, Zhang, and Zhang]{xiang2024far}
J.~Xiang, X.~Xu, F.~Kong, M.~Wu, H.~Zhang, and Y.~Zhang, ``How far can we go with practical function-level program repair?'' \emph{arXiv preprint arXiv:2404.12833}, 2024.

\bibitem[Ramos et~al.(2024)Ramos, Mamede, Jain, Canelas, Gamboa, and Goues]{ramos2024large}
D.~Ramos, C.~Mamede, K.~Jain, P.~Canelas, C.~Gamboa, and C.~L. Goues, ``Are large language models memorizing bug benchmarks?'' \emph{arXiv preprint arXiv:2411.13323}, 2024.

\bibitem[AI(2024)]{codestral}
M.~AI, ``Codestral: Hello, world!'' \url{https://mistral.ai/news/codestral/}, 2024, [Accessed 04-12-2024].

\bibitem[Team et~al.(2024)Team, Zhao, Hui, Howland, Nguyen, Zuo, Hu, Choquette-Choo, Shen, Kelley, et~al.]{team2024codegemma}
C.~Team, H.~Zhao, J.~Hui, J.~Howland, N.~Nguyen, S.~Zuo, A.~Hu, C.~A. Choquette-Choo, J.~Shen, J.~Kelley \emph{et~al.}, ``Codegemma: Open code models based on gemma,'' \emph{arXiv preprint arXiv:2406.11409}, 2024.

\bibitem[Zhang et~al.(2022)Zhang, Panthaplackel, Nie, Li, and Gligoric]{zhang2022coditt5}
J.~Zhang, S.~Panthaplackel, P.~Nie, J.~J. Li, and M.~Gligoric, ``Coditt5: Pretraining for source code and natural language editing,'' in \emph{Proceedings of the 37th IEEE/ACM International Conference on Automated Software Engineering}, 2022, pp. 1--12.

\bibitem[Zirak and Hemmati(2024)]{zirak2022improving}
A.~Zirak and H.~Hemmati, ``Improving automated program repair with domain adaptation,'' \emph{ACM Transactions on Software Engineering and Methodology}, vol.~33, no.~3, pp. 1--43, 2024.

\bibitem[Jin et~al.(2023)Jin, Shahriar, Tufano, Shi, Lu, Sundaresan, and Svyatkovskiy]{jin2023inferfix}
M.~Jin, S.~Shahriar, M.~Tufano, X.~Shi, S.~Lu, N.~Sundaresan, and A.~Svyatkovskiy, ``Inferfix: End-to-end program repair with llms,'' \emph{arXiv preprint arXiv:2303.07263}, 2023.

\bibitem[Paul et~al.(2023)Paul, Hossain, Siddiq, Hasan, Iqbal, and Santos]{paulenhancing}
R.~Paul, M.~M. Hossain, M.~L. Siddiq, M.~Hasan, A.~Iqbal, and J.~Santos, ``Enhancing automated program repair through fine-tuning and prompt engineering,'' \emph{arXiv preprint arXiv:2304.07840}, 2023.

\bibitem[Hao et~al.(2023)Hao, Shi, Liu, and Shu]{hao2023enhancing}
S.~Hao, X.~Shi, H.~Liu, and Y.~Shu, ``Enhancing code language models for program repair by curricular fine-tuning framework,'' in \emph{2023 IEEE International Conference on Software Maintenance and Evolution (ICSME)}.\hskip 1em plus 0.5em minus 0.4em\relax IEEE, 2023, pp. 136--146.

\bibitem[Hossain et~al.(2024)Hossain, Jiang, Zhou, Li, Chiang, Lyu, Nguyen, and Tripp]{Hossain2024}
S.~B. Hossain, N.~Jiang, Q.~Zhou, X.~Li, W.-H. Chiang, Y.~Lyu, H.~Nguyen, and O.~Tripp, ``A deep dive into large language models for automated bug localization and repair,'' \emph{Proceedings of the ACM on Software Engineering}, vol.~1, no. FSE, pp. 1471--1493, 2024.

\bibitem[Yang et~al.(2024{\natexlab{b}})Yang, Tian, Ren, Zhang, Klein, Bissyand{\'e}, Goues, and Jin]{yang2024multi}
B.~Yang, H.~Tian, J.~Ren, H.~Zhang, J.~Klein, T.~F. Bissyand{\'e}, C.~L. Goues, and S.~Jin, ``Multi-objective fine-tuning for enhanced program repair with llms,'' \emph{arXiv preprint arXiv:2404.12636}, 2024.

\bibitem[Ahmed and Devanbu(2023)]{ahmed2023majority}
T.~Ahmed and P.~Devanbu, ``Majority rule: better patching via self-consistency,'' \emph{arXiv preprint arXiv:2306.00108}, 2023.

\bibitem[Huang et~al.(2023{\natexlab{b}})Huang, Zhu, Xing, Jin, Wang, and Xu]{huang2023chain}
Q.~Huang, J.~Zhu, Z.~Xing, H.~Jin, C.~Wang, and X.~Xu, ``A chain of ai-based solutions for resolving fqns and fixing syntax errors in partial code,'' \emph{arXiv preprint arXiv:2306.11981}, 2023.

\bibitem[Deligiannis et~al.(2023)Deligiannis, Lal, Mehrotra, and Rastogi]{deligiannis2023fixing}
P.~Deligiannis, A.~Lal, N.~Mehrotra, and A.~Rastogi, ``Fixing rust compilation errors using llms,'' \emph{arXiv preprint arXiv:2308.05177}, 2023.

\bibitem[{Chen} et~al.(2019){Chen}, {Kommrusch}, {Tufano}, {Pouchet}, {Poshyvanyk}, and {Monperrus}]{SEQUENCER}
Z.~{Chen}, S.~J. {Kommrusch}, M.~{Tufano}, L.~{Pouchet}, D.~{Poshyvanyk}, and M.~{Monperrus}, ``Sequencer: Sequence-to-sequence learning for end-to-end program repair,'' \emph{IEEE Transactions on Software Engineering}, 2019.

\bibitem[Gupta et~al.(2017)Gupta, Pal, Kanade, and Shevade]{deepfix}
R.~Gupta, S.~Pal, A.~Kanade, and S.~Shevade, ``Deepfix: Fixing common c language errors by deep learning,'' in \emph{Proceedings of the Thirty-First AAAI Conference on Artificial Intelligence}, ser. AAAI'17.\hskip 1em plus 0.5em minus 0.4em\relax AAAI Press, 2017, p. 1345–1351.

\bibitem[Abhinav et~al.(2021)Abhinav, Sharvani, Dubey, D’Souza, Bhardwaj, Jain, and Arora]{abhinav2021repairnet}
K.~Abhinav, V.~Sharvani, A.~Dubey, M.~D’Souza, N.~Bhardwaj, S.~Jain, and V.~Arora, ``Repairnet: contextual sequence-to-sequence network for automated program repair,'' in \emph{International Conference on Artificial Intelligence in Education}.\hskip 1em plus 0.5em minus 0.4em\relax Springer, 2021, pp. 3--15.

\bibitem[Chen et~al.(2022)Chen, Kommrusch, and Monperrus]{chen2022neural}
Z.~Chen, S.~Kommrusch, and M.~Monperrus, ``Neural transfer learning for repairing security vulnerabilities in c code,'' \emph{IEEE Transactions on Software Engineering}, vol.~49, no.~1, pp. 147--165, 2022.

\bibitem[Yasunaga and Liang(2020)]{yasunaga2020graph}
M.~Yasunaga and P.~Liang, ``Graph-based, self-supervised program repair from diagnostic feedback,'' in \emph{International Conference on Machine Learning}.\hskip 1em plus 0.5em minus 0.4em\relax PMLR, 2020, pp. 10\,799--10\,808.

\bibitem[Muennighoff et~al.(2023)Muennighoff, Liu, Zebaze, Zheng, Hui, Zhuo, Singh, Tang, von Werra, and Longpre]{muennighoff2023octopack}
N.~Muennighoff, Q.~Liu, A.~Zebaze, Q.~Zheng, B.~Hui, T.~Y. Zhuo, S.~Singh, X.~Tang, L.~von Werra, and S.~Longpre, ``Octopack: Instruction tuning code large language models,'' \emph{arXiv preprint arXiv:2308.07124}, 2023.

\bibitem[Wang et~al.(2023{\natexlab{b}})Wang, Chen, Li, Luo, Peng, Dong, and Liao]{wang2023one}
D.~Wang, B.~Chen, S.~Li, W.~Luo, S.~Peng, W.~Dong, and X.~Liao, ``One adapter for all programming languages? adapter tuning for code search and summarization,'' \emph{arXiv preprint arXiv:2303.15822}, 2023.

\bibitem[Weyssow et~al.(2023)Weyssow, Zhou, Kim, Lo, and Sahraoui]{weyssow2023exploring}
M.~Weyssow, X.~Zhou, K.~Kim, D.~Lo, and H.~Sahraoui, ``Exploring parameter-efficient fine-tuning techniques for code generation with large language models,'' \emph{arXiv preprint arXiv:2308.10462}, 2023.

\bibitem[Wang et~al.(2022)Wang, Yang, Gao, Peng, Zhang, and Lyu]{wang2022no}
C.~Wang, Y.~Yang, C.~Gao, Y.~Peng, H.~Zhang, and M.~R. Lyu, ``No more fine-tuning? an experimental evaluation of prompt tuning in code intelligence,'' in \emph{Proceedings of the 30th ACM Joint European Software Engineering Conference and Symposium on the Foundations of Software Engineering}, 2022, pp. 382--394.

\bibitem[Choi and Lee(2023)]{choi2023codeprompt}
Y.~Choi and J.-H. Lee, ``Codeprompt: Task-agnostic prefix tuning for program and language generation,'' in \emph{Findings of the Association for Computational Linguistics: ACL 2023}, 2023, pp. 5282--5297.

\bibitem[Shi et~al.(2023)Shi, Wang, Zhang, Du, Han, Zhang, and Sun]{shi2023towards}
E.~Shi, Y.~Wang, H.~Zhang, L.~Du, S.~Han, D.~Zhang, and H.~Sun, ``Towards efficient fine-tuning of pre-trained code models: An experimental study and beyond,'' \emph{arXiv preprint arXiv:2304.05216}, 2023.

\bibitem[Zou et~al.(2023)Zou, Li, Ge, Li, Shen, Huang, and Luo]{zou2023comprehensive}
W.~Zou, Q.~Li, J.~Ge, C.~Li, X.~Shen, L.~Huang, and B.~Luo, ``A comprehensive evaluation of parameter-efficient fine-tuning on software engineering tasks,'' \emph{arXiv preprint arXiv:2312.15614}, 2023.

\bibitem[Liu et~al.(2024)Liu, Keung, Yang, Liu, Zhou, and Liao]{liu2024delving}
S.~Liu, J.~Keung, Z.~Yang, F.~Liu, Q.~Zhou, and Y.~Liao, ``Delving into parameter-efficient fine-tuning in code change learning: An empirical study,'' \emph{arXiv preprint arXiv:2402.06247}, 2024.

\end{thebibliography}

\newpage

\section{Biography Section}

\begin{IEEEbiography}[{\includegraphics[width=1in,height=1.25in,clip,keepaspectratio]{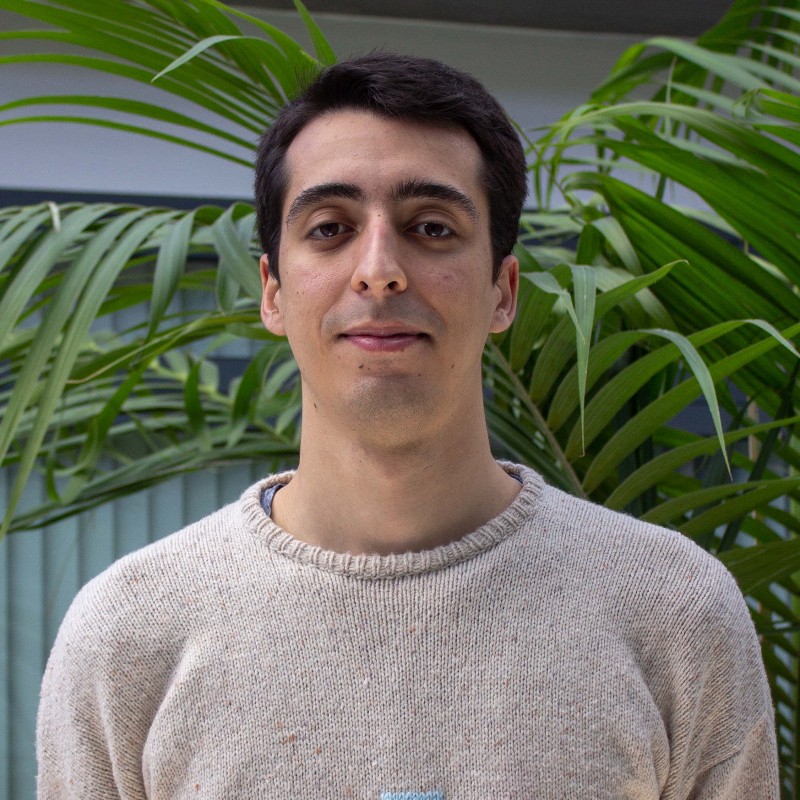}}]{André Silva}
is a Ph.D. student at KTH Royal Institute of Technology, Stockholm, 11428, Sweden. His research interests include the intersection of automatic program repair and machine learning. Silva received his M.Sc. in Computer Science from Instituto Superior Técnico, Universidade de Lisboa, Lisbon, Portugal. Contact him at andreans@kth.se.
\end{IEEEbiography}

\begin{IEEEbiography}[{\includegraphics[width=1in,height=1.25in,clip,keepaspectratio]{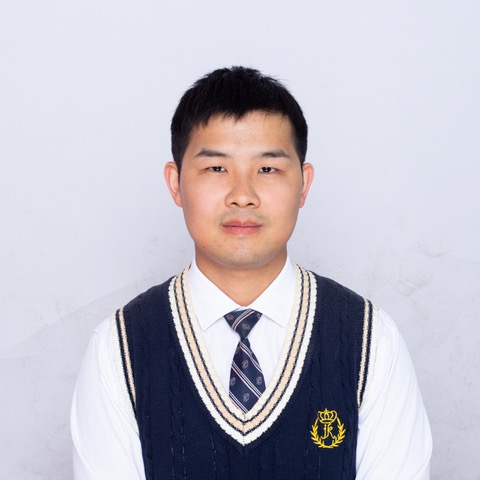}}]{Sen Fang}
is a first-year Ph.D. student at NC State University. His research lies in the intersection between software engineering and machine learning, focusing on AI for software engineering, LLMs for code, and code security. He was the research engineer at KTH, Royal Institute of Techcnology, Sweden.
\end{IEEEbiography}

\begin{IEEEbiography}[{\includegraphics[width=1in,height=1.25in,clip,keepaspectratio]{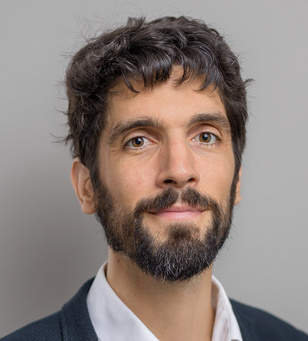}}]{Martin Monperrus}
is Professor of Software Technology at KTH Royal Institute of Technology, Sweden. His research lies in the field of software engineering with a current focus on automatic program repair, AI on code and software integrity. He received a Ph.D. from the University of Rennes, and a Master's degree from Compiègne University of Technology. Homepage: \url{https://www.monperrus.net/martin/}
\end{IEEEbiography}

\end{document}